\documentclass[a4,12pt,twocolumn]{article}
\usepackage{graphicx} 
\usepackage{authblk}
\usepackage[left=1.5cm, right=2cm]{geometry}
\usepackage{amsmath}
\usepackage{tabularx}
\usepackage{booktabs}
\usepackage{hyperref}

\title{Optimization of Coulomb Energies in Gigantic Configurational Spaces of Multi-Element Ionic Crystals}

\author[1,2]{Konstantin Köster}

\author[3]{Tobias Binninger}

\author[1,2]{Payam Kaghazchi\footnote{p.kaghazchi@fz-juelich.de}}

\affil[1]{Materials Synthesis and Processing (IMD-2), Institute of Energy Materials and Devices, Forschungszentrum Jülich GmbH, 52425 Jülich, Germany}

\affil[2]{MESA+ Institute, University of Twente, Hallenweg 15, 7522, NH, Enschede, The Netherlands}

\affil[3]{Theory and Computation of Energy Materials (IET-3), Institute of Energy Technologies, Forschungszentrum Jülich GmbH, 52425, Jülich, Germany}

\date{July 2024}

\begin{document}

\twocolumn[
  \begin{@twocolumnfalse}
    \maketitle
    \begin{abstract}
      Most of the novel energy materials contain multiple elements occupying a single site in their lattice. The exceedingly large configurational space of these materials imposes challenges in determining their ground-state structures. Coulomb energies of possible configurations generally show a satisfactory correlation to computed energies at higher levels of theory and thus allow to screen for minimum-energy structures. Employing a second-order cluster expansion, we obtain an efficient Coulomb energy optimizer using Monte Carlo and Genetic Algorithms. The presented optimization package, GOAC (Global Optimization of Atomistic Configurations by Coulomb), can achieve a speed up of several orders of magnitude compared to existing software. Our code is able to find low-energy configurations of complex systems involving up to 10\textsuperscript{920} structural configurations. The GOAC package thus provides an efficient method for constructing ground-state atomistic models for multi-element materials with gigantic configurational spaces.
    \end{abstract}
  \end{@twocolumnfalse}
]

\section*{Introduction}

Many state-of-the-art solid-state high performance materials are composed of several different types of elements sharing the same lattice sites. Examples for application areas are, but not limited to, energy conversion and storage systems\cite{Chen.2011, Boyd.2019, Anantharamulu.2011, Li.2020, Lun.2021, Sarkar.2018, Fabbri.2010, Yan.2021, Bai.2020, Shin.2015, Wang.2020} as well as other special-purpose applications\cite{Oses.2020, Li.2021, KumarKatiyar.2021, Ma.2021, Yarema.2018}. In some of the most interesting materials for these applications (e.g., layered oxides, ionic conductors), numerous element types with various concentration ratios are combined in a single-crystal phase. While such compositions can be represented with the help of partial site occupations, the configurational complexity becomes a severe challenge for simulation methods that require structural models with integer site occupations, such as commonly used density functional theory~(DFT)\cite{He.2019, Zhang.2013}. The problem of determining reasonable atomistic configurations out of all possible configurations therefore constitutes a serious challenge for modelling and simulation\cite{Wang.2023, dAvezac.2008, Islam.2014, Zhang.2019, TodaCaraballo.2017, Huo.2023}. To represent complex compositions with integer occupations the so-called supercell approach is frequently employed, where multiple periodic images of the unit cell are treated explicitly. For computational studies it is often of interest to determine the lowest-energy atomistic configuration which can be a hard combinatorial problem for large supercells. For complex compositions it is generally infeasible to evaluate all possible configurations (even when accounting for symmetry), especially when using high-level methods such as DFT. Therefore, special techniques such as the Coherent Potential Approximation~(CPA)~\cite{Velicky.1969}, Special Quasirandom Structure~(SQS)\cite{Wei.1990}, Cluster Expansion~(CE)\cite{Laks.1992}, Virtual Crystal Approximation~(VCA)~\cite{Bellaiche.2000}, or Small Set of Ordered Structures~(SSOS)~\cite{Sorkin.2020} have been developed that approximate the energy and/or are able to find special atomistic configurations that have relevant properties for further investigations. Approximations such as CE where many-particle interaction terms up to a certain order are taken into account can reduce the computational demand drastically\cite{Angqvist.2019}. Other approaches that try to mimic highly accurate energies at low computational costs include machine-learned potentials and/or try to reduce the amount of configurations that must be evaluated with other machine learning approaches, e.g., active learning\cite{Kostiuchenko.2019, Yuan.2023, Ferrari.2020, TetsassiFeugmo.2021, Peng.2023, Yaghoobi.2022}.

Naturally, the number of possible configurations becomes higher if the supercell contains more sites, more positions per site, and also when more elements can occupy a site, especially when elements are mixed in equimolar amounts. All of these factors generally apply to novel energy materials and yield a combinatorial explosion of the total number of possible configurations. For highly symmetric cells, this number can be reduced by several orders of magnitude if symmetry operations are taken into account and only symmetrically irreducible configurations are considered\cite{Lian.2022}. There are several software packages and methods such as the site-occupancy disorder~(SOD) code\cite{GrauCrespo.2007}, \textsc{enumlib}\cite{Hart.2008}~(also accessible through \textsc{pymatgen}\cite{Ong.2013}), the solid-solution tools\cite{Mustapha.2013, DArco.2013} in the commercial \textsc{CRYSTAL} code\cite{Erba.2023}, the so-called \textsc{supercell} software\cite{Okhotnikov.2016}, the \textsc{disorder} code\cite{Lian.2020} and its recently published tree search algorithm\cite{Lian.2022}, and the \textsc{SHRY} package\cite{Prayogo.2022} that all focus explicitly on determining symmetry in-equivalent structures. The number of available software and considerable computational effort spent highlights the importance of the atomistic combinatorial problem in computational materials research. 

For ionic crystals, the Coulomb energy with ionic point charges represents a simple energy model allowing to evaluate numerous atomistic configurations with limited computational resources. In practice, the model requires the assignment of the ion valencies and the electrostatic energy is calculated by Ewald summation \cite{Ewald.1921} to obtain the exact Coulomb energy of the periodic lattice\cite{Toukmaji.1996}. This allows to consider plenty of atomistic configurations explicitly and, in some cases, even the complete enumeration of all possible configurations for practical simulation supercells. This full enumeration approach, sometimes also referred to as brute force method or exhaustive sampling, is implemented with Coulomb energy evaluation in the so-called \textsc{supercell} software package\cite{Okhotnikov.2016}. More recently, the \textsc{EwaldSolidSolution} software\cite{Jang.2023} was released offering the brute-force approach with an option for sparser sampling of the density of states based on Coulomb energy evaluation. In addition, \textsc{EwaldSolidSolution} also features a post-processing gradient-descent-like algorithm for optimizing atomistic configurations. However, treating complex combinatorial problems as they appear in modern energy materials by brute forcing is computationally very demanding, even for simple Coulomb energy evaluation. Therefore, classical optimization approaches and the use of heuristics is commonly required.

The atomistic combinatorial sampling can be considered as a general optimization problem and commonly used meta heuristics can be applied. Do~Lee~\textit{et al.}\cite{Lee.2022} applied some well-known heuristics, including genetic algorithms, particle swarm optimization, harmony search, cuckoo search, bayesian optimization, and deep Q-networks, to configurational optimization in argyrodite utilizing Coulomb energies. Out of the vast amount of meta heuristics especially the Genetic Algorithm~(GA)\cite{Fraser.1957} should be mentioned that is know to be effective for the atomistic combinatorial problem\cite{Lee.2022, Han.2018}, as well as for global optimization of complex chemical structures in general\cite{Dieterich.2010}. Next to these classical approaches, more physically motivated approaches such as Monte Carlo~(MC)\cite{Metropolis.1953} simulations were also shown to be efficient in approaching the atomistic configurations problem\cite{Binninger.04.01.2024, Ferrari.2023, Binninger.2020, Kostiuchenko.2019}, with the respective Monte Carlo methods implemented, e.g., for determination of SQS in the \textsc{mcsqs} code\cite{vandeWalle.2013} as part of the \textsc{Alloy Theoretic Automated Toolkit}~(ATAT)\cite{vandeWalle.2002} or for general Cluster expansions in the recently released \textsc{Statistical Mechanics on Lattices package}~\cite{BarrosoLuque.2022}. Binninger~\textit{et al.}\cite{Binninger.04.01.2024} recently also demonstrated that the configuration problem can be solved on existing quantum-computing hardware by formulating it as a binary optimization problem that can be mapped onto a quantum annealer.

The aforementioned software and approaches for determining lowest energy atomistic configurations are either effectively or explicitly limited in the size of the configurational space\cite{Okhotnikov.2016, Lee.2022, Binninger.04.01.2024, Mustapha.2013, DArco.2013, Ong.2013, GrauCrespo.2007} or do not specifically aim to determine the global minimum Coulomb energy structure by optimization\cite{Jang.2023, Hart.2008, vandeWalle.2013, Lian.2020, Lian.2022, BarrosoLuque.2022}. As modern high performance materials introduce more and more species, approaches are required that can reliably and quickly optimize even large combinatorial problems comprising of ten to the power of several hundreds of configurations. For that purpose, either heuristics or general purpose optimization software can be used while the latter one was not applied to the configuration problem by now but bears the opportunity for exact global optimization within limited computational resources. Even though some works already employed heuristic optimization methods to the configuration problem, as discussed before, there is still, to the best of our knowledge, no published tool that allows for optimization of such complex problems yet. Efficient energy evaluation methods, even faster than the commonly applied Ewald summation, along with specifically tailored heuristics must be employed to achieve optimization in difficult atomistic combinatorial problems within reasonable computation time. Creating optimized atomistic configurations for complex problems in a high-throughput manner allows for efficient structure pre-selection for computational studies, such as DFT calculations, of novel materials and thereby offers the opportunity to enhance computational materials discovery in several important research fields.

In this work, we therefore approach the atomistic combinatorial problem in novel energy materials as an optimization problem utilizing a basic but reformulated Coulomb energy model. We present a Python-based code, termed GOAC (Global Optimization of Atomistic Configurations by Coulomb), that enables to interface any configuration problem of ions with distinctive valancies given as a crystallographic information file~(CIF)\cite{Hall.1991} to existing (free or commercial) optimization software. Moreover, we introduce several Fortran-based routines that can be called from the Python code to apply various heuristics to the configurational optimization problem, including GA and MC. To provide a highly efficient implementation, the Coulombic energy is expressed by a second order cluster expansion and the optimization heuristics are parallelized using OpenMP\cite{Dagum.1998}. The methodological details of the implementations and the capabilities of the GOAC code are discussed in the next section, followed by a discussion of the results and benchmarking to alternative methods.

\section*{Methodology}

Structure models in this work were visualize with the \textsc{VESTA} software\cite{Momma.2008}.

\subsection*{Implementation and Theoretical Background}

A supercell is assumed comprising $S$ sites with partial occupations and each site having $P_s$ positions within the cell. Moreover, a site should be occupied by $N_{s,e}$ ions of the element $e$ while in total $E_s$ elements can occupy the given site $s$. The total number of possible configurations $C$ in the supercell, without considering any symmetries, is then given by:

\begin{equation}
    C = \prod_{s=1}^S \frac{P_s!}{\prod_{e=1}^{E_s}N_{s,e}!}.\label{eq:combinations}
\end{equation}

\begin{figure}[tb]
\centering
\includegraphics[width=0.475\textwidth]{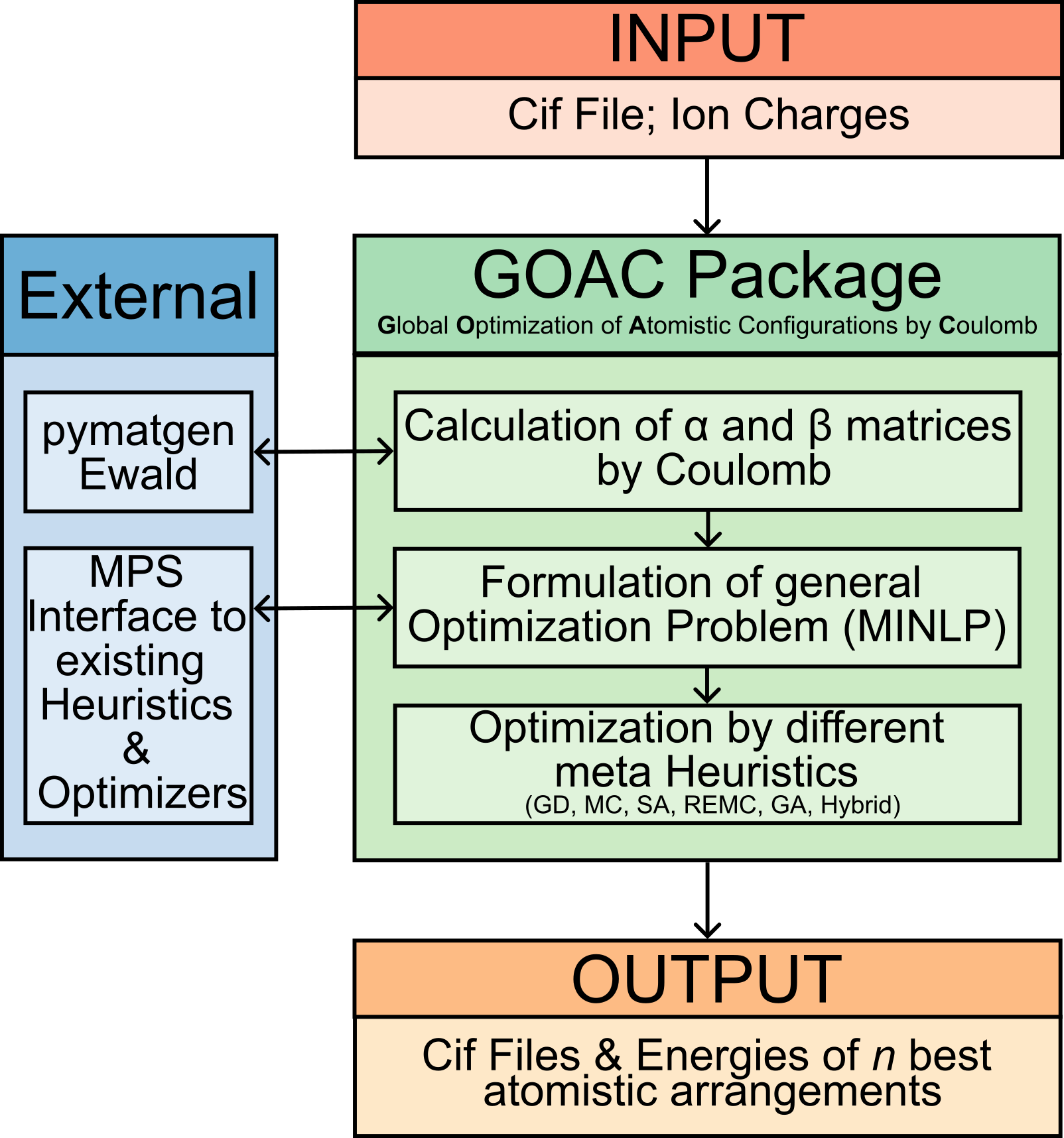}
\caption{Schematic workflow of the GOAC code and connection to external packages.}\label{fig:GOAC}
\end{figure}

For a given problem, the Global Optimization of Atomistic Configurations by Coulomb~(GOAC) code aims to determine low(est) energy atomistic configuration(s) out of all possible configurations employing various optimization techniques. To this end, GOAC offers a command line interface to provide a CIF with partial occupations and assumed charge states (valencies) for the different ions. The general workflow of GOAC is sketched in Figure~\ref{fig:GOAC}.

In a first step, GOAC calculates the required energy matrices for a cluster expansion of the ionic Coulomb energy, which is discussed in the next section. Then, an optimization problem is constructed that can be either interfaced to external optimizers, e.g., the \textsc{Gurobi} solver\cite{GurobiOptimization.2023}, or solved by internal Fortran heuristics. Both approaches are discussed in the following sections. Finally, the $n$ lowest energy atomistic configurations are outputted as a CIF along with the respective Coulomb energies. It should be noted that, in its current implementation, GOAC is not able to identify symmetry-equivalent structures and all optimizers run on the full configurational space.

\subsubsection*{Cluster Expansion of Coulomb Energy}

\begin{figure}[tb]
\centering
\includegraphics[width=0.475\textwidth]{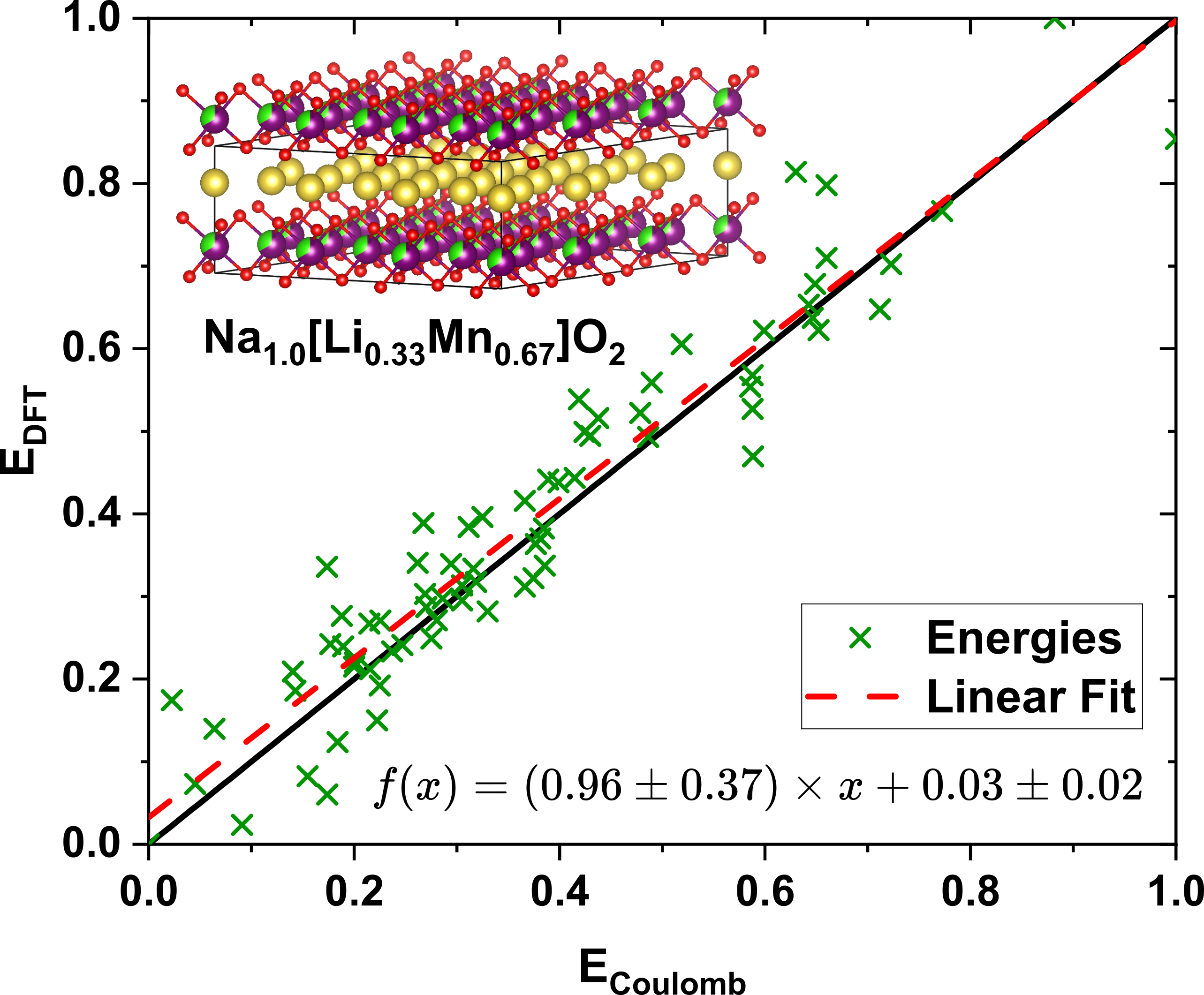}
\caption{Correlation between normalized relative DFT (details are described in the Method section) and Coulomb energies of different ionic configurations for Na[Li\textsubscript{0.33}Mn\textsubscript{0.67}]O\textsubscript{2}. Coulomb energies were obtained with the following ionic valencies: Na: +1; Li: +1; Mn: +4; O: $-$2. A linear fit of the data points is shown as a red dashed line, along with the ideal correlation diagonal (black solid line).}\label{fig:DFT}
\end{figure}

As optimization methods generally require evaluating the energy of many different atomistic configurations, GOAC implements an ionic Coulomb energy model due to the low computational demand. Naturally, such simple point charge models cannot account for quantum mechanical effects and there is no guarantee that the order of different ionic configurations by Coulomb energy is aligned with the one obtained by more accurate calculations, e.g., based on DFT. However, several studies showed that structures with a low Coulomb energy are often also good candidates for low DFT energies\cite{Binninger.2020, Jang.2023, Binninger.04.01.2024, Lee.2022, Moradabadi.2019}. As an example, a satisfactory correlation between DFT and Coulomb energies is shown in Figure~\ref{fig:DFT} for ionic configurations in the layered oxide Na[Li\textsubscript{0.33}Mn\textsubscript{0.67}]O\textsubscript{2} (assumed ionic charges: Na: +1; Li: +1; Mn: +4; O: $-$2). The normalized relative energies show a strong correlation between DFT and Coulomb models and the linear fit well matches the diagonal representing perfect correlation. A commonly employed approach therefore consists in pre-selecting a certain number of low Coulomb energy structures to be used for more accurate DFT calculations and eventually determine low DFT energy configurations\cite{Moradabadi.2019, Kim.2020, Voronina.2023, Kim.2024, Pahari.2023}.

The DFT reference calculations shown in Figure~\ref{fig:DFT} were performed with the \textsc{Vienna ab initio Simulation Package}~(VASP)\cite{Kresse.1996} in the projector augmented wave~(PAW) scheme\cite{Blochl.1994} with the Perdew-Burke-Ernzerhof~(PBE) exchange-correlation functional\cite{Perdew.1996}. An energy cut-off of 520\,eV along with a convergence criterion of $10^{-4}$\,eV, a 1$\times$1$\times$2 $\Gamma$-centered $k$-point grid, and spin-polarization was employed. Single-point calculations without any geometry optimization were performed to allow for a fair comparison to Coulomb energies. The exact geometries can be found in the ``Examples'' folder of the project repository.

Following this approach, GOAC utilizes point-charge Coulomb energies in a second-order cluster expansions model, meaning that cluster interactions up to the second order (particle-particle interactions) are included in the total energy calculation. We note that for the specific case of the point-charge Coulomb energy the second-order CE is exact due to the pairwise character of Coulomb point-charge interactions. This allows for an efficient evaluation of different atomistic configurations during optimization as the energy can be expressed as a sum of pre-calculated coefficients.

\begin{figure*}[tb]
\centering
\includegraphics[width=0.9\textwidth]{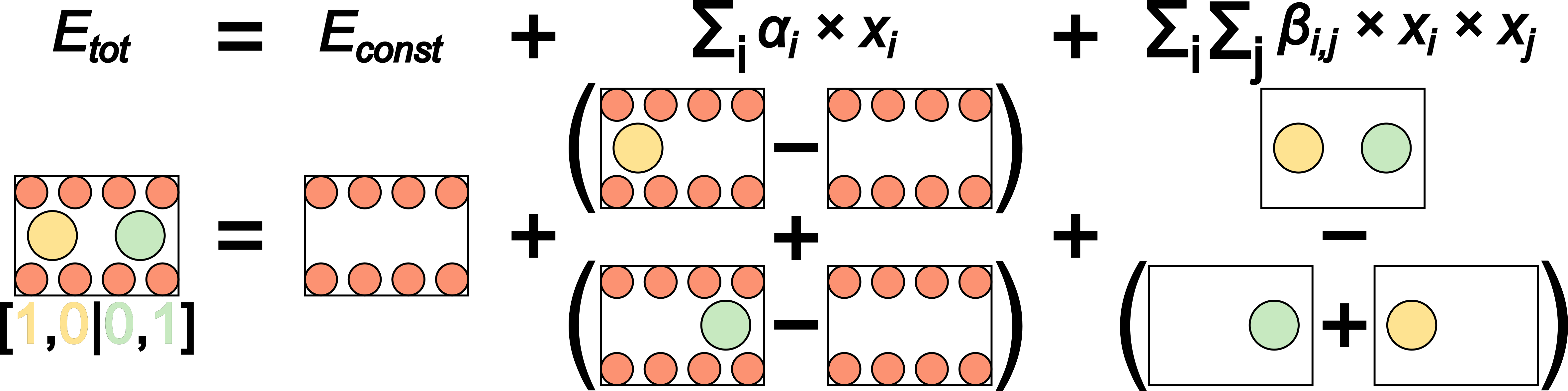}
\caption{Schematic visualization of the CE approach for the energy calculation of atomistic configurations along with the simplified energy formula and an example on how to map specific atomistic configurations on a binary vector.}\label{fig:CE}
\end{figure*}

The procedure of expressing the atomistic combinatorial problem in a CE is sketched in Figure~\ref{fig:CE}. The total energy~($E_{tot}$) of a given atomistic configuration can be expressed as a sum of the energy of the fixed ions (zero-order term, $E_{const}$), the interaction of each placed iterative ion with the fixed ions as well as its self-interaction due to periodic boundary conditions (first-order term, $\alpha$), and all particle-particle interactions between all placed iterative sites (second-order term, $\beta$). To avoid double-counting, the first-order term has to be corrected for the interactions that are already considered in the zero-order term and the second-order term must be corrected for the self-interactions that are already included in the first-order term. Energies for the sub-structures that are considered in the expansion can be pre-calculated which allows for efficient evaluation of different atomistic configurations during optimization. For the exemplary problem in Figure~\ref{fig:CE} with two sites that are both occupied by 50~\% by two different species, all possible configurations can be expressed by a binary solution vector $x$ that has a position for each site for each species. A possible solution would than have a 1 on every position where a species is placed and a 0 everywhere else. By that, the total energy of a given instance becomes a simple sum of products of pre-calculated first-order~($\alpha$) and second-order~($\beta$) coefficients and the binary solution vector $x$. To ensure that only second-order terms are counted where both ions are placed, the $\beta$-coefficients are multiplied by the two corresponding positions in the binary solution vector.

By default, the energy is evaluated in terms of Coulomb energy in GOAC. In that case, a second order cluster expansion is even exact due to the fact that only particle-particle interactions are considered in Coulomb. In periodic systems, Coulomb energies are, however, difficult to converge and the Ewald summation technique is required for the calculation of the respective CE coefficients. The corresponding sub-structures are not charge-balanced but Ewald summation requires charge-neutrality\cite{Hub.2014}. However, these errors are perfectly eliminated in the CE approach as the total charge of the considered sub-systems add up to the charge of the complete system, i.e., zero for an overall charge-neutral system. The CE approach, as shown in the equation in Figure~\ref{fig:CE}, is therefore capable to calculate exact periodic Coulomb energies from the pre-calculated terms. For the implementation in GOAC, the \textsc{pymatgen}\cite{Ong.2013} package is leveraged to calculate the CE coefficients in terms of the Ewald energies of the respective sub-structures.

For implementing the CE a slight reformulation of the CE appears to be practical where the solution vector $x$ has one more dimension, one for the site-species~($i$) and one for the positions this site-species can occupy~($j$). Consequently, the cluster expansion coefficients $\alpha$ and $\beta$ become higher in dimensionality as well. By reformulation of the sums it is ensured that each interaction is only counted in one direction and just one half of the diagonal $\alpha$ and $\beta$ matrices must be stored. Lastly, for a full optimization problem the constraints have to be defined. Beyond the binary constraint for the $x$ variables~(Equation~\ref{eq:Opt-Problem-c3}) it must be also ensured by additional constraints that the desired total occupancy~($O_i$) is matched for each site-species~$i$~(Equation~\ref{eq:Opt-Problem-c1}) and that a certain position $j$ is not occupied by multiple species $i$~(Equation~\ref{eq:Opt-Problem-c2}). In summary, the optimization problem of atomistic configurations is implemented in GOAC as shown in Equation~\ref{eq:Opt-Problem}.

\begin{subequations}\label{eq:Opt-Problem}
    \begin{align}
    &\min_{E_{tot}} \quad E_{tot} = E_{const} + \sum_{i=1}^{S}\sum_{j=1}^{P_i} \alpha_{i,j} \cdot x_{i,j} + \tag{\ref{eq:Opt-Problem}}\\
    &\sum_{i=1}^{S}\sum_{j=1}^{P_i}\sum_{l=j+1}^{P_i} \beta_{i,j,i,l} \cdot (x_{i,j} \cdot x_{i,l}) + \nonumber\\ 
    &\sum_{i=1}^{S}\sum_{j=1}^{P_i} \sum_{k=i+1}^{S}\sum_{l=1}^{P_k} \beta_{i,j,k,l} \cdot (x_{i,j} \cdot x_{k,l}) \nonumber\\
    &\text{subject to: } \nonumber\\
    &\sum_{j=1}^{P_i} x_{i,j} = O_i \quad \ \forall i \in S \label{eq:Opt-Problem-c1}\\
    &\sum_{i=1}^{S} x_{i,j} \leq 1 \quad\quad \forall j \in P_i  \label{eq:Opt-Problem-c2}\\
    &x_{i,j} \in \{0;1\} \quad\quad \forall i \in S; \quad \forall j \in P_i \label{eq:Opt-Problem-c3}
    \end{align}
\end{subequations}

Even though Coulomb (Ewald summation) calculations are computationally comparably inexpensive, for high-throughput evaluations of atomistic configurations Equation~\ref{eq:Opt-Problem} represents a significant speed-up compared to a full Ewald summation for each atomistic configuration. GOAC also supports multi-threading for the energy calculations of the sub-structures. Moreover, by storing the CE coefficients, the pre-calculated energy terms conveniently allow to test multiple optimization approaches without performing extensive energy calculations every time. It should be noted that it is possible to write all sub-structures that appear in the second order CE as CIFs and evaluate the structures by other, external methods, if needed. The resulting energies can be fed back into GOAC to make use of the optimizers described in the following section.

\subsection*{Optimization Strategies for Atomistic Configurations}\label{sec:meth:opt}

Two main categories of optimizers, namely exact and heuristic optimizers, can be distinguished. A successful run of an exact optimizer guarantees that the global optimum is found or, if specified, not just the global optimum but the $n$ lowest energy structures while $n$ can be freely chosen by the user. The heuristic optimizers guarantee to output a valid, low energy structure that might be the global minimum or just a local minimum or no minimum at all, depending on the optimizer. The focus of heuristics is to create valid, high-quality solutions fast, while exact optimizers spend significant effort on proving optimality without improving the actual minimum solution. Depending on the needs of the user, both approaches can be valuable and are accessible via the GOAC code as described in the next sections.

\subsubsection*{Interfacing to External Exact Optimizers}

Generally speaking, Equation~\ref{eq:Opt-Problem} describes a so-called mixed integer non-linear programming~(MINLP) problem with the special circumstance that all variables are not just integer but binary variables which technically allows for a reformulation to a mixed integer linear programming~(MILP) problem. Problems of the same type frequently appear the context of business economics under the collective term \textit{Operations Research}, where the aim is, e.g., to determine the optimal~(shortest/fastest) delivery route\cite{Zhang.2022} or to optimize production planning\cite{Shamsaei.2017}. Due to the economic value connected to this problem type plenty of optimizers exist\cite{Kronqvist.2019}. Their aim is to find the global optimum and also prove that the global optimum was found employing advanced mathematical strategies that can be faster than a full enumeration of all possible solutions~(brute forcing), which, by definition, is also an exact optimization method. 

For a given atomistic combinatorial problem, GOAC can create a standard MINLP with the help of the licensed Gurobi\cite{GurobiOptimization.2023} software and the full problem statement is written to a standard MPS~(Mathematical Programming System) file. By default, GOAC passes this MINLP also to Gurobi for solving, however, it should be noted that the MPS file can be used to run the problem in other (commercial or free) optimization software. GOAC supports interfacing to the Gurobi optimizer and its solver parameters. It is worth noting that Gurobi (and other software) is technically capable of linearizing the quadratic terms in the MINLP to an MILP due to the binary character of the integer variables. This is not done by default in GOAC but was found to be efficient for some problems. Such a reformulation can also allow the use of other standard optimization software that are not capable of general MINLPs. However, results for exact optimizations presented in this work were obtained with the default Gurobi parameter set in GOAC, which was found to be most robust for different configuration problems. It should be noted that the MPS file of the problem can be also handed to non-exact heuristic solvers.

\subsubsection*{Internal Fortran Heuristics in GOAC}

The core of the GOAC code offers different heuristic optimizers for the atomistic combinatorial problem that are all tailored for this specific problem and implemented in Fortran. All of these heuristics are capable of generating valid low energy structures. The following methods are currently supported in the GOAC code: a random structure generator, a Greedy Heuristic, a Gradient Descent algorithm~(GD), a Metropolis Monte Carlo code~(MC)\cite{Metropolis.1953}, a simulated annealing extension of the MC code~(SA), a Replica Exchange Monte Carlo scheme~(REMC)\cite{Thachuk.2007}, and a Genetic Algorithm~(GA)\cite{Fraser.1957} with roulette wheel selection\cite{Lipowski.2012}. Due to the possibility of accessing GOAC as a python package it is also straight forward to combine some of the aforementioned heuristics to a hybrid approach. Such combinations were already proposed and proven successful for chemical optimization problems\cite{Dugan.2009, Sakae.2015} and a combination of the REMC and GA heuristic is benchmarked and referred to as HY in the following. The implementations of the different algorithms are discussed in more detail in the supplementary information and the code can be directly accessed within the project repository. 

Most heuristics that are directly implemented in the GOAC code are of stochastic nature and it can be useful to run the same heuristic multiple times. By that procedure, the probability and confidence that the global minimum is found can be increased. For convenience, GOAC allows to run the same heuristic multiple times in parallel with the help of OpenMP\cite{Dagum.1998} which allows to achieve an statistic ensemble over multiple runs with the same heuristic. Moreover, trivial parallelizations such as, e.g., parallelization over the different temperatures in REMC are also implemented via OpenMP in GOAC to further boost the performance of the code. The scaling behavior of the different algorithms is also sketched in Supplementary~Figure~1. Finally, the internal heuristics in GOAC offer abortion by run time or heuristic steps without improvement on the global minimum. More detailed descriptions of GOAC's features and how to employ them can be found in the manual inside the project repository.

\section*{Resulst and Discussion}
\subsection*{Performance of Exact Optimization Methods}\label{sec:res:exact}

As explained above, GOAC has the possibility to interface to external optimization software for exact optimization of atomistic configurations. For this benchmark, the Gurobi optimizer, which utilizes an advanced branch-and-cut method, is employed with the default parameter set GOAC uses to interface to Gurobi. This parameter set enforces strong pre-solving of the model (Presolve=2) along with a focus on proving optimality (MIPFocus=2). It also ensures that the $n$ lowest energy structures are found by setting the convergence boundaries to zero (MIPGap=0 and MIPGapAbs=0) and the ``PoolSearchMode'' to 2. To the best of our knowledge, the existing software for exact optimization of configurations, i.e., including proof of optimality, employ the full enumeration approach. An efficient implementation of the latter can be found in the \textsc{supercell} software, which is used as a reference for timings of full enumeration. Here it should be noted that the \textsc{supercell} software only considers the symmetry in-equivalent structures which reduces the number of explicitly considered atomistic configurations drastically compared to the total number of configurations when ignoring symmetry. 

\begin{figure}[tb]
\centering
\includegraphics[width=0.475\textwidth]{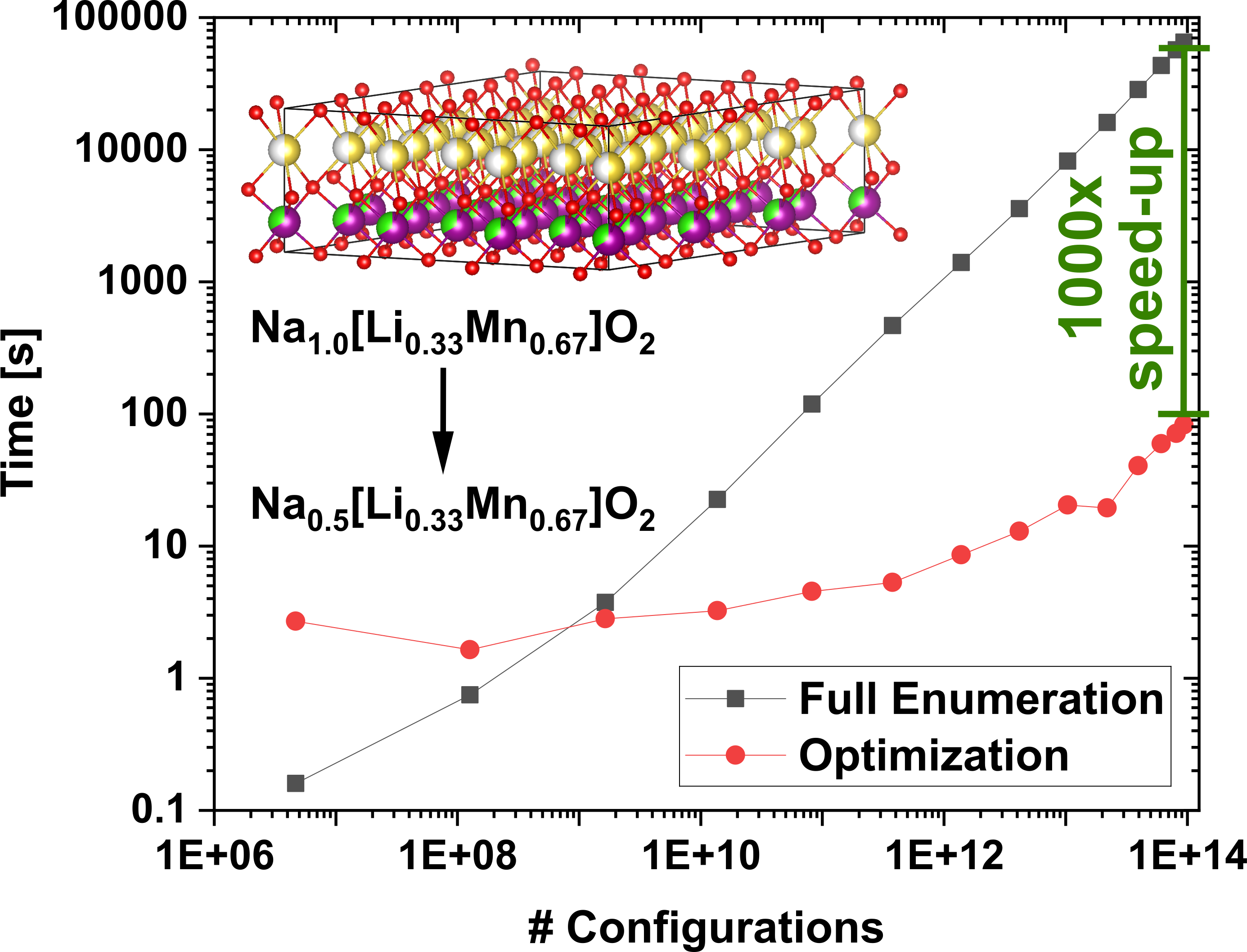}
\caption{$\log_{10}$-$\log_{10}$ representation of run time (estimated from outputted timings of the software) to find the global optimum atomistic configuration versus number of total possible configurations using the full enumeration approach and external optimization software.}\label{fig:Exact-Opt}
\end{figure}

The \textsc{supercell} software and the optimization with Gurobi of the model prepared by GOAC were tested on a layered-oxide sodium-ion-battery cathode material~(Na[Li\textsubscript{1/3}Mn\textsubscript{2/3}]O\textsubscript{2})\cite{Wang.2021} with one layer in the \textit{c}-direction and partial occupations in both the transition-metal and sodium-ion sites, cf. Figure~\ref{fig:Exact-Opt}. By changing the sodium-ion stoichiometry from 1.0 to 0.52, configuration combinatorics with steadily increasing number of total possible configurations ranging from ca. $10^7$ to $10^{14}$ were created and evaluated by both approaches. Charge states of Na\textsuperscript{+}, Li\textsuperscript{+}, Mn\textsuperscript{4+} and a variable oxidation state of oxygen ranging from $-$2.0 (for a Na stoichiometry of 1) to $-$1.75926 (for a Na stoichiometry of 0.52) were assumed as the compound is reported to show anionic redox\cite{Wang.2021}. As both approaches guarantee to determine the global optimum after a successful run, it is only of interest to benchmark the run time of both methods. The timings on 128 physical processor cores are plotted against the total number of configurations in Figure~\ref{fig:Exact-Opt}. For smaller problem instances with up to ca. $10^9$ configurations, full enumeration was faster than optimization due to the overhead of interfacing to an external optimization code combined with the capability of the \textsc{supercell} software to reduce the solution space to just symmetry in-equivalent structures. However, it should be noted that timings on these problem instances are well below 10 seconds and therefore computationally inexpensive in both approaches. For more complex problems the full enumeration approach scales perfectly linearly while run time of the branch-and-cut optimization method increased more irregularly from the small offset caused by the overhead. In general, the computation time of the branch-and-cut optimization was significantly lower for more complex instances of this problem and also appeared to scale lower towards problems with many configurations. Overall, a speed-up of up to three orders of magnitude was achieved by the optimization with Gurobi compared to full enumeration with the \textsc{supercell} software at the most difficult considered problem instance with ca. $10^{14}$ total configurations. The respective run times to find the global optimum atomistic configuration in Coulomb energy were ca. 18 hours by full enumeration versus ca. 1.5 minutes by Gurobi optimization. 

Figure~\ref{fig:Exact-Opt} clearly highlights the computational advantages that can be accessed by using GOAC to formulate a general optimization problem for the combinatorial ground-state search that can be handed to external optimization software. However, extrapolating the scaling behavior to much larger problems also reveals that even with the significant speed up achieved, still only problems of intermediate difficulty/size can be tackled. It must be also noted that the actual performance of the branch-and-cut optimization is very much problem dependent. By introducing (slight) changes to the presented problem (e.g., more species per site or more sodium sites by using a P-type layered structure\cite{Delmas.1980}), problems can be constructed where optimization of the complete configuration space is even slower than full enumeration of symmetry in-equivalent structures or problems that formally have as many as $10^{230}$ configurations, but are being optimized within a second, might be obtained.

In summary, the performance of applying standard optimization software to the atomistic combinatorial problem is strongly problem (material) dependent. However, our results indicate that especially for problems of intermediate difficulty (ca. $10^{10}$ to $10^{20}$ possible configurations), such as configuration of charge carriers in rechargeable energy storage materials, optimization can give a significant computational advantage over full enumeration approaches, even if the full enumeration method accounts for symmetry equivalents.

\subsection*{Benchmark of Heuristics in GOAC}\label{sec:res:heu}

As the heuristics do not guarantee to find the global minimum, a suitable benchmark could either compare the lowest energy that is found within given computational resources or the time that it takes to find a known global optimum. However, the implemented heuristics are of stochastic nature which makes it important to average their performance over multiple runs. Such comparisons of the different internal heuristics in GOAC are discussed for several examples with various complexity in the following. Moreover, an additional benchmark of FeSbO\textsubscript{4} is shown in the supplementary information. All examples in the following were executed on the same hardware and run times (given in real time) were estimated by the CPU time required to perform each calculation.

\subsubsection*{Atomistic Configurations in NaCl}

The site occupation in NaCl is not a true combinatorics problem as the unit cell contains two distinctive sites, one for Na and one for Cl. However, for testing purposes both sites can be modified such that each site is occupied by 50~\% of each species, yielding an atomistic combinatorial problem. With this model, in a 3$\times$3$\times$3-supercell the total number of possible configurations is ca. $10^{64}$, a rather difficult combinatorial problem. As the global optimum still remains trivial, a perfectly alternating pattern of Na and Cl in all dimensions, this problem statement is a rather suitable benchmark. Moreover, calculation of the Madelung constant\cite{Madelung.1918},
\begin{equation}
    M_C = \frac{4 \pi \times \epsilon_0 \times r \times |E|}{N_{\text{Ions}}/2 \times e} \ , \label{eq:madelung}
\end{equation}
is straight forward and convergence to the literature value of $M=1.74756...$\cite{Sakamoto.1958} can be tracked for the different heuristics over run time. In this equation, $\epsilon_0$ is the electric constant, $r$ the lattice distance of two neighbouring sites (2.81\,\AA), $E$ the Coulomb energy of the considered structure, $N_{\text{ions}}$ the total number of ions in the structure (216), and $e$ the elementary charge.

\begin{figure*}[tb]
\centering
\includegraphics[width=0.96\textwidth]{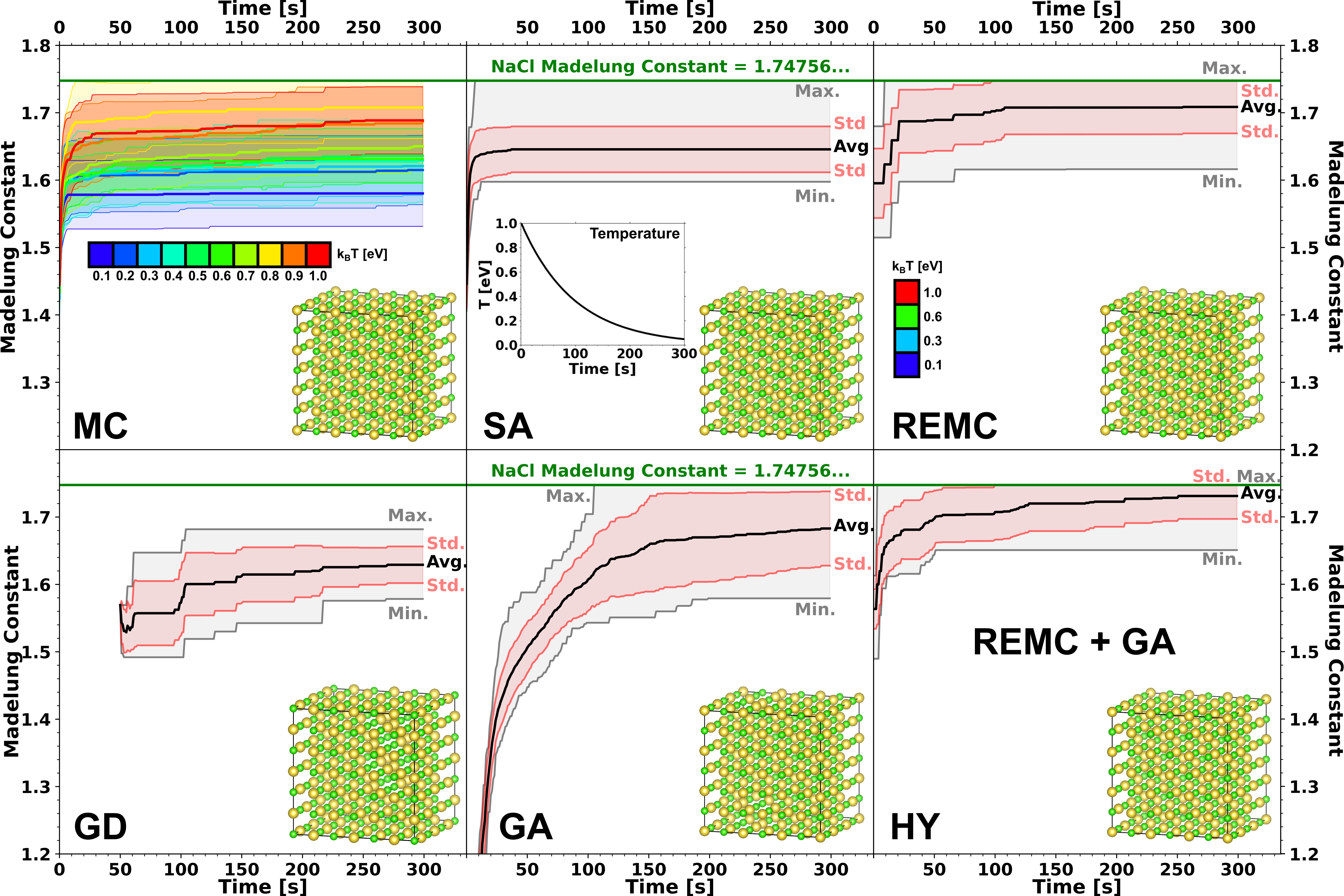}
\caption{Single-Core (1 physical CPU core, 1 OpenMP thread) performance of the MC, SA, REMC, GD, GA, and HY approaches as implemented in GOAC on the atomistic combinatorial problem of NaCl in a 3$\times$3$\times$3-supercell with $10^{64}$ possible configurations. All heuristics were run for 300 seconds and averages over 16 independent runs along with their statistics (standard deviations, minimum, maximum) and the obtained minimum structure are shown in the plots. For MC, averages and standard deviations over 16 runs are visualized for 10 different temperatures, respectively. The temperature evolution for SA is shown in the respective inset and the REMC parallel temperatures are indicated in the corresponding plot.}\label{fig:NaCl}
\end{figure*}

The convergence towards the Madelung constant for the heuristics implemented in GOAC is plotted in Figure~\ref{fig:NaCl}. It is observed that the Gradient Descent heuristic requires some time before the first solutions can be obtained. In this algorithm, the first solution is written as soon as the local minimization from a random starting point is finished and then the next random starting structure is optimized. The time required to reach this first solution is also different for random starting structures as different amounts of optimization steps are necessary to reach a local minimum. Therefore, in the beginning of the GD plot, averages over less than 16 runs are contained, which also explains the drop in the average caused by more independent runs that obtained their first solution being included. Even though the average becomes flatter and standard deviations as well as min-max differences become smaller towards the end of the 5 minutes run time, no run was able to find the global minimum. This highlights the problem of this algorithm as it guarantees to find a local minimum but on shallow energy surfaces with many local minima it becomes highly unlikely to find the global minimum as there is a high chance to get trapped in another local minimum.

The same tendency can be observed in Figure~\ref{fig:NaCl} for Monte Carlo performed at low temperatures (ca. 0.1--0.5\,eV) as the average quickly flattens to a constant value since the algorithm gets trapped in local minima at a low sampling temperature, similar to the outcome of the GD method. At higher temperatures (ca. 0.6--0.8\,eV), however, the averages are observed to get closer to the Madelung constant (corresponding to the global energy minimum) over time as local barriers can be passed with a certain probability to eventually find lower minima. At high temperatures (ca. 0.9--1.0\,eV) the algorithm is able to pass even higher energy barriers, thus spending only short times for local optimization and resulting in a decrease of average performance. In this example, the best result (on average) was obtained at a temperature of 0.8\,eV and the performance was quite sensitive to the simulation temperature, even though multiple runs at various temperatures were able to find the global optimum within five minutes of run time.

The average performance of Simulated Annealing was similar to that of MC at lower to intermediate temperatures with a relatively high variance in solutions, as some runs returned the global optimum. This behaviour can be explained by the rather fast cooling rate chosen, which exponentially decreased from an initial simulation temperature of 1.0\,eV to almost 0\,eV during the run time (cf. inset in Figure~\ref{fig:NaCl}). Such a high cooling rate, which was required to scan a sufficiently large temperature range within the given run time limit, makes it more unlikely that a sufficient temperature is present at the crucial optimization steps leading to a high risk of local minima trapping. Nevertheless, SA was able to find the global optimum in some runs.

The last tested approach from the MC family, namely Replica Exchange Monte Carlo, shows a better performance than SA. The algorithm showed a pronounced optimization, especially in the first ca. 100 seconds, before almost constant values for average, standard deviation, and min-max were reached. This behaviour indicates that the optimization got trapped in local minima for some runs, while in other runs the global optimum was successfully reached. As only about one third of the run time (ca. the first 100 seconds) was effectively used for optimization, the performance might be improved by using more than four temperatures in REMC, including also largely-different and higher temperatures. Compared to the other heuristics, REMC performed very well within the given run time.

Among the approaches compared in Figure~\ref{fig:NaCl}, the Genetic Algorithm shows the slowest increase in average performance versus run time. Several generations and selection procedures are required to obtain more optimized structures resulting in the steep improvement of average energy. Even though some GA runs successfully reached the global optimum, the average over all runs was still substantially  below the correct Madelung constant after 300 seconds of run time, showing that some runs got trapped in local minima. The trapping also goes along with high standard deviations and a large min-max difference. This occurs if the structural variation in the generations becomes low and centered around a deep local minimum. Another problem can be that the generation consists of symmetry equivalents of the same local minimum or if the local minimum is so deep that it can not be exited at small mutation rates which are required for a systematic optimization.

To overcome these limitations the Hybrid approach can be employed which provided the best performance among the methods compared in Figure~\ref{fig:NaCl}. Here, a pre-trained (from REMC) generation was used for the GA which greatly improved the average performance within the first seconds of the run. Moreover, the REMC steps between the GA runs can help to improve the variation in the generation pool of the GA. Vice versa, the GA offers a systematic procedure to make rather large steps on the potential energy surface that cannot be efficiently achieved by pure REMC. Therefore, both approaches can complement each other and the results demonstrate that HY was very effective with the average of 16 independent runs being fairly close to the correct Madelung constant after just 5 minutes of run time and with many runs ending in the global optimum. Moreover, the average kept increasing at longer run times indicating that most of the runs would eventually converge to the global optimum. Notably, the HY strategy performed better than the two individual approaches (GA and REMC) and was the best out of all investigated methods, indicating that a beneficial synergy effect between GA and REMC was achieved.

\subsubsection*{Li-Site Occupation and Ta Doping in LLZO}

Li\textsubscript{7}La\textsubscript{3}Zr\textsubscript{2}O\textsubscript{12}~(LLZO) is a widely studied electrolyte for all-solid-state batteries and therefore of high practical interest. However, the global minimum energy structure is rather hard to approach computationally due to its large unit cell (8 formula units). The computational challenge becomes even more severe when dopants and defects are introduced that require even larger supercells. For these cases, the configurational space is extremely large, representing an interesting test for GOAC to obtain optimized atomistic configurations in terms of Coulomb energies. As an example, we consider Li\textsubscript{6}La\textsubscript{2.969}Zr\textsubscript{0.906}Ta\textsubscript{1.094}O\textsubscript{12} (Charges: Li\textsuperscript{1+}, La\textsuperscript{3+}, Zr\textsuperscript{4+}, Ta\textsuperscript{5+}, O\textsuperscript{2$-$}) which can be modelled by a 2$\times$2$\times$1 supercell (32 formula units). The modelled composition is in good agreement with the experimental one reported by Redhammer~\textit{et al.}\cite{Redhammer.2021}. We define the structure such that all lithium ions can be placed in both the tetrahedral and octahedral sites, resulting in a total of ca. $10^{159}$ possible atomistic configurations. The corresponding structure model is also shown in Supplementary~Figure~2. 

\begin{figure}[tb]
\centering
\includegraphics[width=0.475\textwidth]{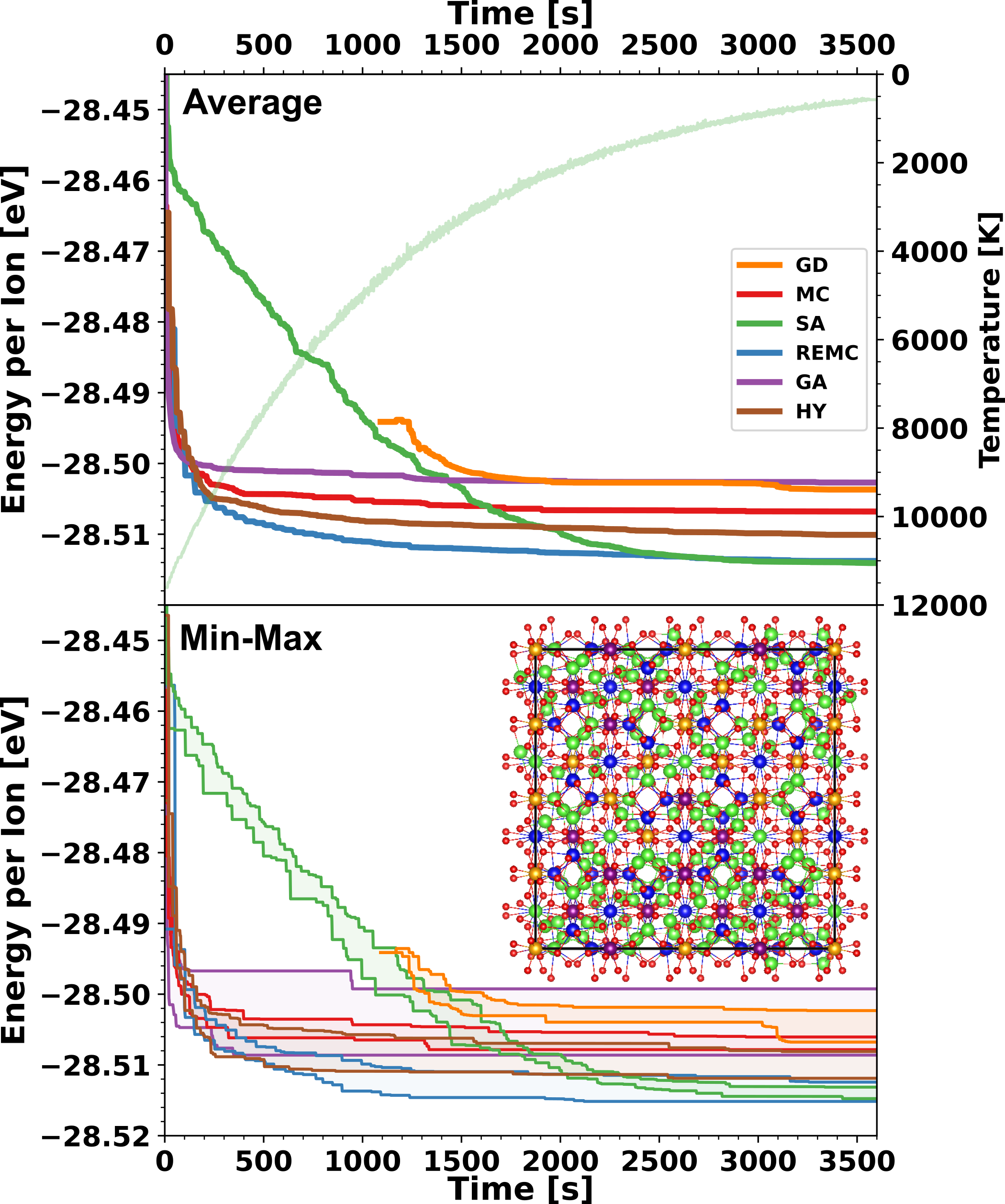}
\caption{Average (upper plot) and min/max (lower plot) energies per ion for the LLZO atomistic combinatorial problem with ca. $10^{159}$ configurations over 10 independent runs of 1 hour at 128 physical CPU cores (128 OpenMP threads) for GD, MC, SA, REMC, GA, and HY as implemented in GOAC. In the upper plot, also the average temperature profile used for the SA simulations is shown. The structure inset in the lower plot corresponds to the minimum energy structure that was obtained across all optimizations. An enlarged version of the minimum energy structure is also given in Supplementary~Figure~3.}\label{fig:LLZO}
\end{figure}

Performances over 10 independent optimization runs are visualized for each heuristic of GOAC in Figure~\ref{fig:LLZO}. As discussed previously at the example of NaCl, the GD algorithm requires some time before the first local optimizations are finished and therefore the average plot begins at ca. 1000 seconds in Figure~\ref{fig:LLZO}. The overall performance of GD was found to be among the worst out of the GOAC heuristics. The GA solutions converged to a similar average energy as GD, but also had the largest variation between the best and worst independent runs, hinting at local minima trapping. This behavior might be reasoned by the different parallelization approaches as discussed in the supplementary information and the code documentation in the project repository. However, averages shown in Figure~\ref{fig:LLZO} are still a fair comparison of optimization performance versus CPU time, revealing that the heuristics including some sort of MC are more efficient than a pure GA for LLZO.

The MC approach returned an intermediate average energy per ion, while the SA and REMC methods yielded significantly lower energies after one hour of run time. For most heuristics the convergence was rather flat beyond the first ca. 500 seconds, but SA showed an exponential decrease over the whole run time which was matching the exponential decrease of the respective simulation temperature from ca. 12000\,K to almost 0\,K (cf. Figure~\ref{fig:LLZO}). Interestingly, also the variance between the best and worst runs became rather small for the SA approach. In contrast to the results obtained for NaCl, SA performed well for the present example due to the longer run times that allowed for a slower cooling rate. The final average energies obtained from SA and REMC were similar, but the energy of the best REMC run was slightly lower than that of the best SA run, and the corresponding minimum-energy structure is shown in Figure~\ref{fig:LLZO}. The superior performances of SA and REMC over the other methods demonstrate that MC approaches with some temperature variation are very effective for the complex LLZO configuration problem. While in this example, the HY approach was not able to improve on the performance of REMC, still a much lower average energy than for the pure GA was found. The overall performance of HY might be increased by longer run times and adjusted heuristic parameters.

The determined minimum energy structure can be analysed in terms of the ratio of lithium ions in tetrahedral versus octahedral coordination of oxygen as all lithium ions were freely iterated over both classes of sites during optimization. A ratio of $\frac{77}{115} \approx 0.67$ is obtained which is in very good agreement with the ratios of 0.74, 0.64, and 0.59 (after different treatments) and an average of 0.66 reported from experiments\cite{Redhammer.2021}. This highlights again the predictive quality of point-charge Coulomb energies for the configuration of ions in complex structures and validates the approach of pre-selecting atomistic configurations by Coulomb energies for higher-level calculations. Moreover, in the case of Ta-doped LLZO discussed here, the structure is so complex and the required supercell is so large that it can hardly be approached by any other model or method. To the best of our knowledge, explicit optimization has never been reported before for any comparably complex atomistic combinatorial problem by Coulomb energies. The largest reported example in the literature comprised of ca. $10^{75}$ total configurations\cite{Jang.2023} (a benchmark with GOAC is also shown in Supplementary~Table~1), which is significantly smaller than the LLZO problem discussed here.

\subsubsection*{High-Entropy Layered Oxides}

To further demonstrate the optimization capabilities of GOAC, we addressed the atomistic combinatorial problem in a high-entropy layered sodium-ion-battery cathode material. The composition of O3-Na\textsubscript{2/3}[Li\textsubscript{1/6}Fe\textsubscript{1/6}Co\textsubscript{1/6}Ni\textsubscript{1/6}Mn\textsubscript{1/3}]O\textsubscript{2} was recently proposed by Yao~\textit{et al.}\cite{Yao.2022}, while O3 indicates that the structure has three layers in the $c$-direction and octahedral coordination of the sodium ions\cite{Delmas.1980}. We modeled the system in a $\sqrt{3}$-unit cell ($a=5.0$\,\AA, $c=19.2$\,\AA) assuming ionic charges of Na\textsuperscript{+}, Li\textsuperscript{+}, Fe\textsuperscript{2.5+}, Co\textsuperscript{3.5+}, Ni\textsuperscript{2+}, Mn\textsuperscript{4+}, and O\textsuperscript{1.75$-$}. All sodium ions were iterated over all sodium positions in every layer (one sodium site in the whole structure with nine positions in the unit cell) and all ions in the transition metal layers were iterated over all positions in each layer (one transition metal site in the whole structure with nine positions in the unit cell), allowing for the maximal configurational space. To highlight the scalability of GOAC, this configuration problem was solved in supercells of different sizes ranging from 4 unit cells (2$\times$2$\times$1, Na\textsubscript{24}[Li\textsubscript{6}Fe\textsubscript{6}Co\textsubscript{6}Ni\textsubscript{6}Mn\textsubscript{12}]O\textsubscript{72}) to 108 unit cells (6$\times$6$\times$3, Na\textsubscript{648}[Li\textsubscript{162}Fe\textsubscript{162}Co\textsubscript{162}Ni\textsubscript{162}Mn\textsubscript{324}]O\textsubscript{1944}). Structure models for the smallest and largest considered supercells are visualized in Supplementary~Figure~4. Results for optimizing the atomistic configurations with the heuristics in GOAC within a given run time (given computational resources) are summarized in Table~\ref{tab:Na-layered}.

\begin{table*}[tb]
\centering
\caption{Energies per ion of the lowest energy structures obtained for differently sized supercells of O3-Na\textsubscript{2/3}[Li\textsubscript{1/6}Fe\textsubscript{1/6}Co\textsubscript{1/6}Ni\textsubscript{1/6}Mn\textsubscript{1/3}]O\textsubscript{2} with the different heuristics implemented in GOAC (all calculations performed on 128 physical CPU cores and using 128 OpenMP threads). Even though energy differences appear to be small as they are scaled per ion to allow for comparisons, they are very significant (usually more than 1\,eV) when the energy of a whole supercell is considered. The table also shows the total number of configurations~(\textit{N\textsubscript{Conf}}) of the different combinatorial problems along with the required time to calculate the CE Coulomb matrices~(\textit{t\textsubscript{C}}). The run times for each heuristic were limited to \textit{t\textsubscript{R}} for the shown problems. Minimum energies found for each of the problem sizes are highlighted in bold.}\label{tab:Na-layered}%
\begin{tabular}{@{}cccccccccc@{}}
\toprule
& & & & \multicolumn{6}{@{}c@{}}{Energy per ion [eV]}\\
\cmidrule{5-10}%
$\sqrt{3}$-Cell & \textit{N\textsubscript{Conf}} & \textit{t\textsubscript{C}} [min] & \textit{t\textsubscript{R}} [h] & GD & MC & SA & REMC & GA & HY\\
\midrule
2$\times$2$\times$1 & $10^{31}$ & 0.2167 & 1 & \textbf{-22.080} & \textbf{-22.080} & \textbf{-22.080} & \textbf{-22.080} & \textbf{-22.080} & \textbf{-22.080}\\
2$\times$2$\times$2 & $10^{64}$ & 0.5000 & 1 & -22.071 & -22.082 & \textbf{-22.096} & \textbf{-22.096} & \textbf{-22.096} & \textbf{-22.096}\\
4$\times$4$\times$1 & $10^{132}$ & 1.817 & 1 & -22.040 & -22.027 & \textbf{-22.087} & \textbf{-22.087} & -22.042 & \textbf{-22.087}\\
4$\times$4$\times$2 & $10^{269}$ & 12.57 & 2 & -22.013 & -22.006 & \textbf{-22.096} & \textbf{-22.096} & -21.772 & -22.040\\
6$\times$6$\times$1 & $10^{303}$ & 19.05 & 2 & - & -22.003 & -22.078 & \textbf{-22.083} & -21.814 & -22.029\\
8$\times$8$\times$1 & $10^{543}$ & 134.5 & 4 & - & -21.975 & \textbf{-22.031} & -22.022 & -21.625 & -21.986\\
6$\times$6$\times$2 & $10^{611}$ & 201.6 & 8 & - & -21.984 & \textbf{-22.040} & -22.031 & -21.589 & -21.992\\
6$\times$6$\times$3 & $10^{920}$ & 790.9 & 16 & - & -21.970 & \textbf{-22.018} & -22.006 & -21.548 & -21.975\\
\bottomrule
\end{tabular}
\end{table*}

Remarkably, all solvers were capable to find the same minimum, likely the global minimum, for the smallest problem of a 2$\times$2$\times$1 supercell within just one hour of run time. It should be also noted that most heuristics identified this minimum within the first minutes (cf. the convergence versus run time plots in Supplementary~Figures~5--10). Compared to the exact solvers presented in the previous section, this represents a huge speed up as a problem with $10^{31}$ total configurations would be (almost) impossible to solve with an exact solver in a reasonable run time, especially not within just one hour. This highlights the practicability of GOAC as problems of this size regularly appear when high-entropy structures or similarly complex structures are to be pre-selected for DFT calculations. 

For the next larger problem, a 2$\times$2$\times$2 supercell, only the more advanced heuristics, namely SA, REMC, GA, and HY, were able to find the same lowest energy structure, which makes it again a likely candidate for the global minimum in Coulomb energy. The respective minimum energy is lower than the minimum energy obtained for the smaller problem, because the increased problem size allows for larger, energetically more favourable superstructures. The same applies to the 4$\times$4$\times$1 supercell where SA, REMC, and HY obtained the same best candidate configuration for the global minimum. As the periodicity is extended in a different direction compared to the 2$\times$2$\times$2 supercell, the minimum energy is still lower than for the 2$\times$2$\times$1 case but higher than for the 2$\times$2$\times$2 supercell. For a 4$\times$4$\times$2 supercell, only SA and REMC were capable to find a likely candidate for the global minimum. The respective minimum energy is identical to the one of the 2$\times$2$\times$2 problem as both consider the same periodicity, and thus same degrees of freedom, in the $c$-direction. The additional degrees of freedom in $a$ and $b$-direction, on the other hand, did not seem to allow for the formation of lower energy superstructures. These findings highlight another aspect why it is important to consider sufficiently large supercells in the construction of structural models with occupational disorder, because suitable supercell sizes are required for lowest energy superstructures. 

For an even larger 6$\times$6$\times$1 supercell, the GD heuristic was not able to reach any local minimum within the given run time since more complex problems not only increase the expected number of optimization steps required to reach a local minimum from a random starting structure but also heavily increase the amount of neighbouring structures that need to be evaluated to follow the steepest descent path. Within the given framework, $10^{269}$ configurations seemed to be the maximum where GD could be applied within reasonable computational resources, which is arguably already a quite large configurational space. For the 6$\times$6$\times$1 supercell, REMC returned the lowest energy structure, lower in energy than the 2$\times$2$\times$1 minimum, which was expected given that the 6$\times$6$\times$1 is a multiple of the 2$\times$2$\times$1 supercell. SA also returned a low-energy solution, albeit not the same minimum, probably because the cooling rate was too fast for the given problem size and run time limitation.

For all supercells larger than 6$\times$6$\times$1, SA found the lowest energy structure out of all heuristics implemented in GOAC. However, the obtained minima did  not correspond to the respective global minima as they were higher in energy than the minimum energy structures of one of the smaller problems with matching multiplicity. Still, it is remarkable that both SA and REMC were able to identify high-quality low-energy atomistic configurations within relatively short calculation times for problems with a configurational space up to $10^{920}$, a number even larger than the estimated total number of atoms in the entire universe \cite{Ade.2016} to the power of ten.\footnote{The actual number of atoms in the universe must be estimated from measured densities and hydrogen/helium distributions and is in the range of ca. $10^{80}$ atoms.}

The pure MC heuristic performed inferior to the more elaborate SA and REMC extensions for all problem sizes. As it was shown for NaCl, the MC method is quite sensitive to the simulation temperature which was not re-optimized for every problem in the benchmark (fixed to 0.75\,eV). The GA performed rather poor for problems with a complexity of $10^{269}$ or more in its current implementation. Combining the GA with REMC in the HY approach did not resolve this issue for the larger problem sizes as the gain in performance compared to the pure GA stemmed almost exclusively from the REMC part. Therefore, the overall performance of the HY method was still inferior to using all computational resources on REMC. More advanced HY combination schemes or different crossing strategies in the GA might resolve this under-performance in the future.

\section*{Conclusion}

In this work, we showed that the problem of finding lowest energy configurations in the huge configurational space of modern energy materials can be effectively approached by using advanced optimization methods in combination with Coulomb energy models. The Coulomb energy variations between different configurations often align well with energies from higher levels of theory, e.g., DFT, and sampling by Coulomb energies is therefore an attractive method to pre-select low-energy structure candidates. As a tool for conveniently and effectively exploring the vast configurational space of atomistic configurations in complex materials, we introduced the GOAC code that can be accessed via the command line or as a Python package.

The calculation of energies of different configurations was significantly sped up by applying a cluster expansion to the Coulomb energy cost function, which enables the use of pre-calculated coefficients in the optimization procedure, thus providing significant improvements over alternative approaches that perform Ewald summations at each optimization step. This reformulation transforms the atomistic combinatorial problem statement into an MINLP problem and allows to employ various advanced optimization methods. We showed that the exact optimization of the MINLP, interfaced via GOAC to existing optimization software, was several orders of magnitude faster than the full enumeration approach often applied for the atomistic combinatorial problem, allowing to exactly solve configuration problems for system sizes that could not be approached before.

Due to the combinatorial explosion of the configurational space in complex multi-element materials, exact solving strategies cannot be applied to more complex materials. For such problems, we implemented several heuristics in GOAC, including Gradient Descent, Monte Carlo, Simulated Annealing, Replica Exchange Monte Carlo, Genetic Algorithms, and hybrid approaches. With these heuristics, GOAC produced high-quality low-energy structures within limited computational resources for extremely large configuration problems, which is of interest to model complex compositions and identify possible superstructures. As a highlight, we showed that GOAC was able to find likely candidates for global minimum structures of problems with $10^{269}$ configurations in just about 2 hours of run time on 128 CPU cores. Moreover, it was demonstrated that for problem combinatorics up to $10^{920}$, it was still possible to find good low-energy solutions using the GOAC package. To the best of our knowledge, explicit optimization of atomistic configurations by Coulomb energy has not been reported for such large problems before.

In the current implementation, GOAC is working with simple point-charge Coulomb energies, which represent a rough estimation that does not guarantee to coincide in the lowest energy configuration with higher level of theory approaches, e.g., DFT. Moreover, atomistic combinatorial problems with charge-neutral ions (atoms) or ions with identical valencies cannot be optimized on the basis of Coulomb energies alone. However, GOAC offers the possibility to define the coefficient matrices of the cluster expansion energy model by other approaches which might help to overcome some of these limitations in future works. In summary, GOAC can be a valuable tool for computational research on novel energy materials and other complex materials to determine likely candidate structures for low or lowest energy atomistic configurations with comparably little computational resources.

\section*{Acknowledgements}

The presented work was carried out within the framework of the Helmholtz Association’s program Materials and Technologies for the Energy Transition, Topic 2: Electrochemical Energy Storage. Computation time granted through JARA-HPC on the supercomputer JURECA\cite{Thornig.2021} at Forschungszentrum Jülich under Grant No. jiek12 is gratefully acknowledged by the authors. K.K. and P.K. thank for the financial support from the ``Deutsche Forschungsgemeinschaft'' (DFG, German Research Foundation) under project No. 501562980.

\clearpage
\newpage


\clearpage
\newpage

\renewcommand{\figurename}{Supplementary Figure}
\renewcommand{\tablename}{Supplementary Table}

\twocolumn[
  \begin{@twocolumnfalse}
\section*{Supplementary Information: Optimization of Coulomb Energies in Gigantic Configurational Spaces of Multi-Element Ionic Crystals}
  \end{@twocolumnfalse}
]

\subsection*{Descriptions of the Implementations for the Heuristics in GOAC}

\subsubsection*{Random~(R)} A Fortran routine that randomly selects a site-species and position and then places the corresponding ion if it is allowed by the occupation constraints of the atomistic combinatorial problem. The routine continues this procedure iteratively until a complete valid solution is achieved (e.g, all occupancy constraints are fully matched).

\subsubsection*{Greedy Heuristic~(GH)} The greedy heuristic is a common inexpensive benchmark that can be the upper bond when comparing other methods or which can be used as a starting point for more advanced optimization techniques. Sites with partial occupations are filled iteratively while always the species and position with the lowest energy out of all possibilities is selected to place an ion. The Fortran implementation in GOAC also supports nested selection where the algorithm continues not just with the lowest but the $n$-lowest structures in each iteration. To avoid combinatorial explosion, the pool size is always reduced to $n$ structures after two steps such that at most $n^2$ structures are considered per step.

\subsubsection*{Gradient Descent~(GD)} This Fortran routine performs a local minimization for a given (usually random) starting solution. For a starting solution all neighbouring solutions that can be accessed by swapping one iterative ion with an vacancy or two iterative ions with each other are evaluated and the lowest energy structure is selected until no direct neighbour structure with lower energy exists. This method can therefore be considered as a Gradient Descent for the discrete atomistic combinatorial problem and guarantees to find a local minimum. The routine in GOAC also supports to continue with the $n$ lowest energy neighbours in each iteration while the solution pool is always reduced to $n$ solutions after two steps such that at most $n^2$ solutions are considered per step. The routine that explores all neighbouring solutions can also be considered as a single step of a branch and bond approach and can technically be directly accessed via GOAC to obtain the $n$ lowest energy neighbour structures for a given starting solution. This can help to develop new heuristics with the help of GOAC.

\subsubsection*{Monte Carlo~(MC)} GOAC contains a Fortran implementation of an Metropolis Monte Carlo simulation\cite{Metropolis.1953} for the atomistic combinatorial problem. Again, a random valid starting solution is generated and a random neighbouring solution is chosen. A neighbouring structure is a structure that can be accessed by a single atom-swap. Then, the Metropolis criteria ($P=\min[1, \exp(-\Delta E/kT)]$, with $\Delta E$ being the energy difference of the current structure and the selected neighbour, $k$ the Boltzmann constant, and $T$ the simulation temperature) is used to determine the probability at which the random selected neighbouring structure is accepted. The new structure is then accepted or rejected by the calculated probability and the whole procedure is repeated for a desired amount of simulation steps. 

\subsubsection*{Simulated annealing~(SA)} This routine builds on the aforementioned MC routine but in addition, the temperature can be slowly reduced to 0\,K during the simulation run. With a high starting temperature and a sufficiently slow cooling rate the chance of finding minimum energy structures can be increased.

\subsubsection*{Replica Exchange Monte Carlo~(REMC)} The REMC method also builds upon the aforementioned MC Fortran routine but in this approach multiple MC runs at different temperatures are performed in parallel. After a certain amount of MC simulation steps per temperature, structures are interchanged between the independent simulations. The probability at which an interchange between simulations at different temperatures is accepted is \cite{Thachuk.2007}: $\min[1, \exp(\Delta E \cdot (1/kT_1 - 1/kT_2 ))]$. Such a parallel tempering strategy can help to overcome energy barriers by using higher temperatures but still enables local optimizations in the parallel runs at lower temperatures.

\subsubsection*{Genetic Algorithm~(GA)} A Fortran implementation of a genetic algorithm\cite{Fraser.1957} for the atomistic combinatorial problem which starts from a pool of random structures. Parent structures are selected via roulette wheel method\cite{Lipowski.2012} and the structures are crossed by randomly selecting atom-swaps that are different in the parent structures with an expectation value of accepting 50\% of all possible atom-swaps that constitute the differences in the parent structures. Moreover, random mutation can be applied to the generated structures at a given rate. Furthermore, elitism is supported by directly transferring a certain amount of best structures from the parent pool to the child pool of the next generation (self-reproduction). The algorithm can be run for a desired amount of steps (generations).

\subsubsection*{Hybrid Methods~(HY)} As the aforementioned algorithms can get trapped in local minima which prevents them from reaching the global optimum, it can be beneficial to combine different methods to a hybrid heuristic approach. Especially the GA can run into a local minimum and if too little variation is present in the generation, the algorithm gets trapped. In such cases, it can help to perform local minimization of the generation structures or to apply more dynamic optimization approaches to overcome local barriers and thereby improve the variations in the generation structures.\cite{Gregurick.1996} Combinations of a GA with MC, SA, and REMC were all proposed and proven to be effective for chemical optimization problems.\cite{Dugan.2009, Sakae.2015} In GOAC, such hybrid heuristics can be conveniently created by calling the different Fortran heuristic methods sequentially from a Python script. In this study, a hybrid heuristic with GA and REMC, where the REMC and GA methods are called alternately and the best solutions are passed on, is discussed.

\subsubsection*{External Heuristics~(EH)} GOAC also offers the possibility to interface the optimization problem as MPS file to other heuristic optimization software. GOAC allows for direct interfacing to the heuristics implemented in Gurobi\cite{GurobiOptimization.2023}. A default set of Gurobi parameters is used by GOAC, but a wrapper can be utilized to tweak all Gurobi parameters.

\section*{Scalability of the GOAC Functionalities}
\begin{figure}[h!tb]
\centering
\includegraphics[width=0.475\textwidth]{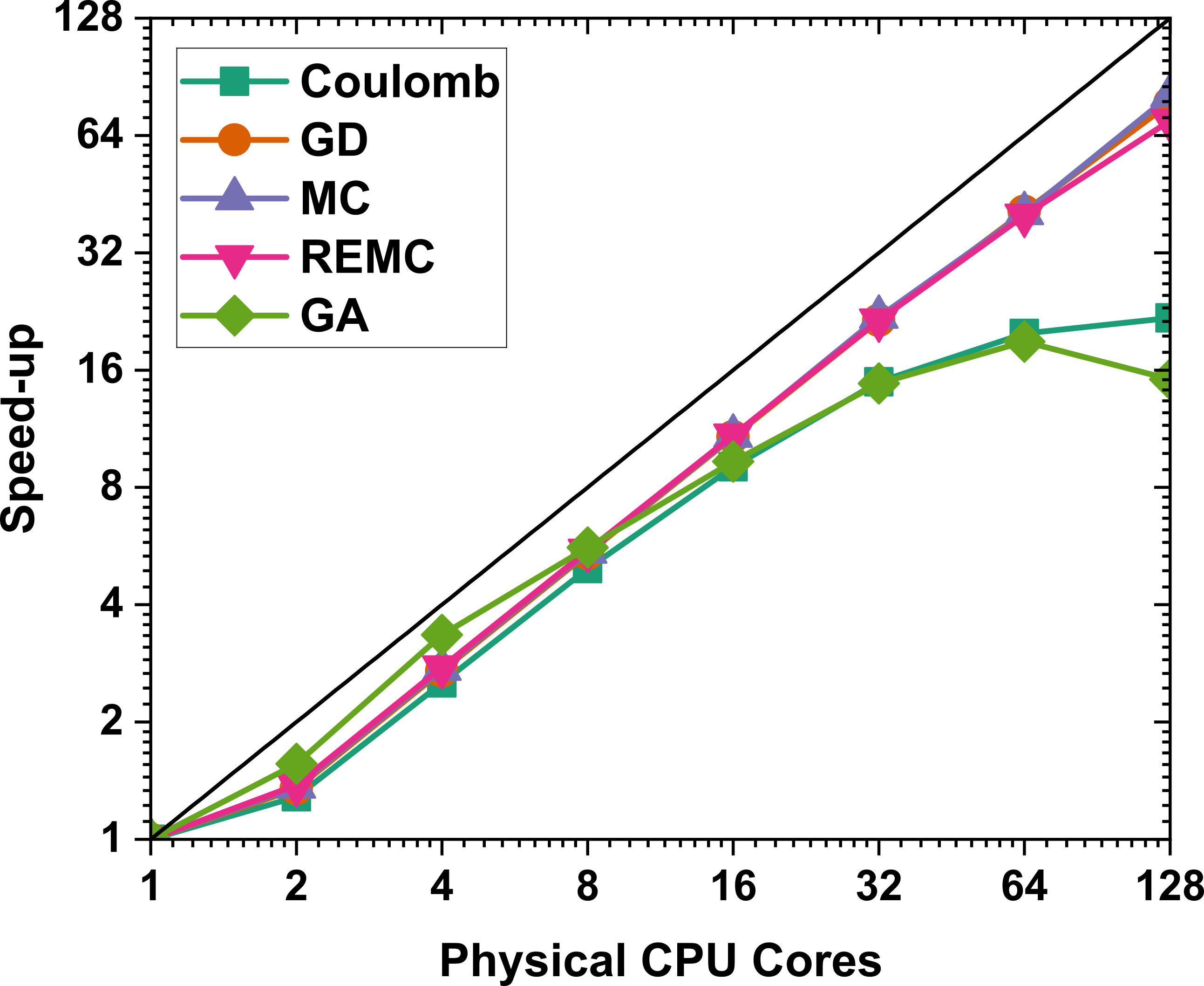}
\caption{$\log_2-\log_2$-plot of the speed-up versus the number of CPU cores (OpenMP threads) for the different parallelized methods in the GOAC code. Ideal scaling behavior is indicated by the black diagonal.}\label{fig:Scaling}
\end{figure}

The scalability plot in Figure~\ref{fig:Scaling} shows that the internal parallelizations for GD, MC, and REMC work rather well. This is also to expect for GD and MC as the parallelization simply consist of running multiple runs in parallel which is effectively the same as running the code multiple times manually. For REMC, however, it proves that also the internal parallelizations over the simulation temperatures are implemented effectively. For the Coulomb calculation (CE coefficient matrices calculations) the scaling becomes slightly worse at many cores and the same applies to the GA where the generation creation is parallelized. These lower scalabilities probably stem from the fact that in these methods parallelization is less trivial than just running the same code multiple times. Moreover, the problems used for this benchmark might have been too simple such that the actual scaling behavior was masked by some overhead. Nevertheless, all methods show a suitable scaling behavior to be effectively employed on machines with multiple cores.

\subsection*{Structure Models of Ta-doped LLZO}

\begin{figure*}[h!tb]
\centering
\includegraphics[width=0.95\textwidth]{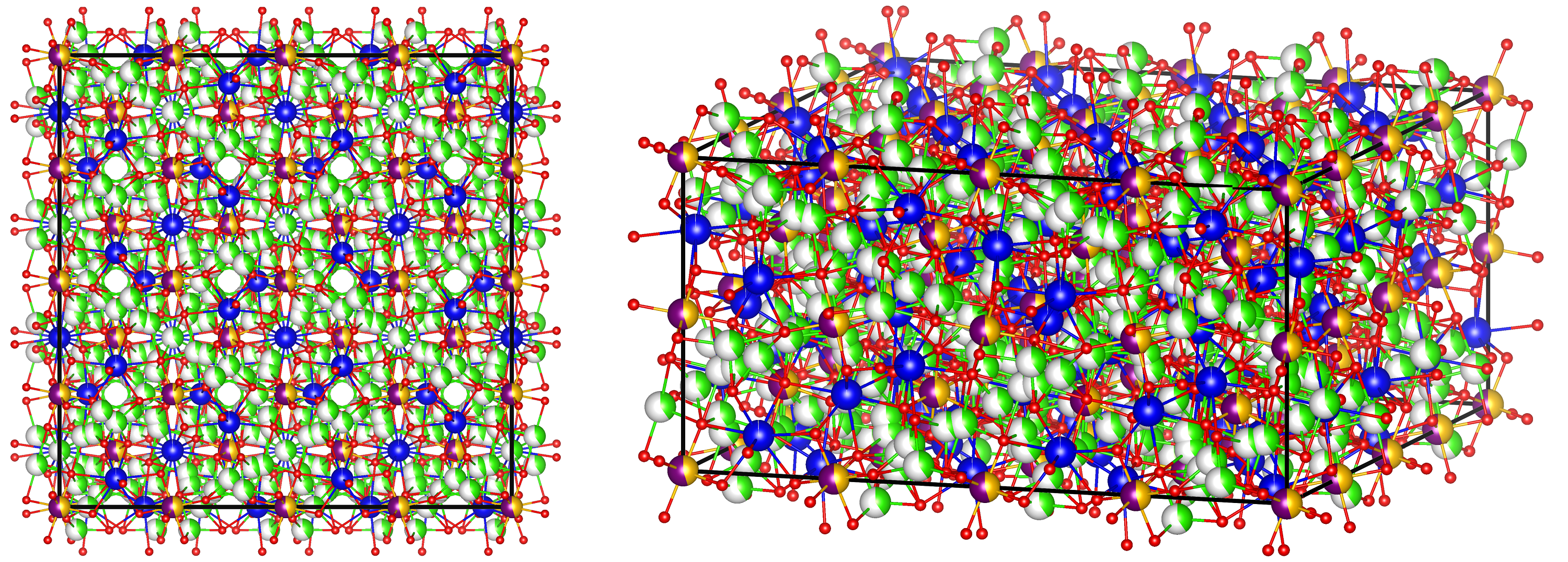}
\caption{Input structure with partial occupations for the optimization of the Ta-doped LLZO presented in the main text in the front (left) and perspective (right) view. A single-site was defined for the Li positions such that all positions shown here were freely iterated for Li. The color-coding is as follows: Li: green, La: blue, Ta: purple, Zr: orange, O: red.}\label{fig:LLZO1}
\end{figure*}

\begin{figure*}[h!tb]
\centering
\includegraphics[width=0.95\textwidth]{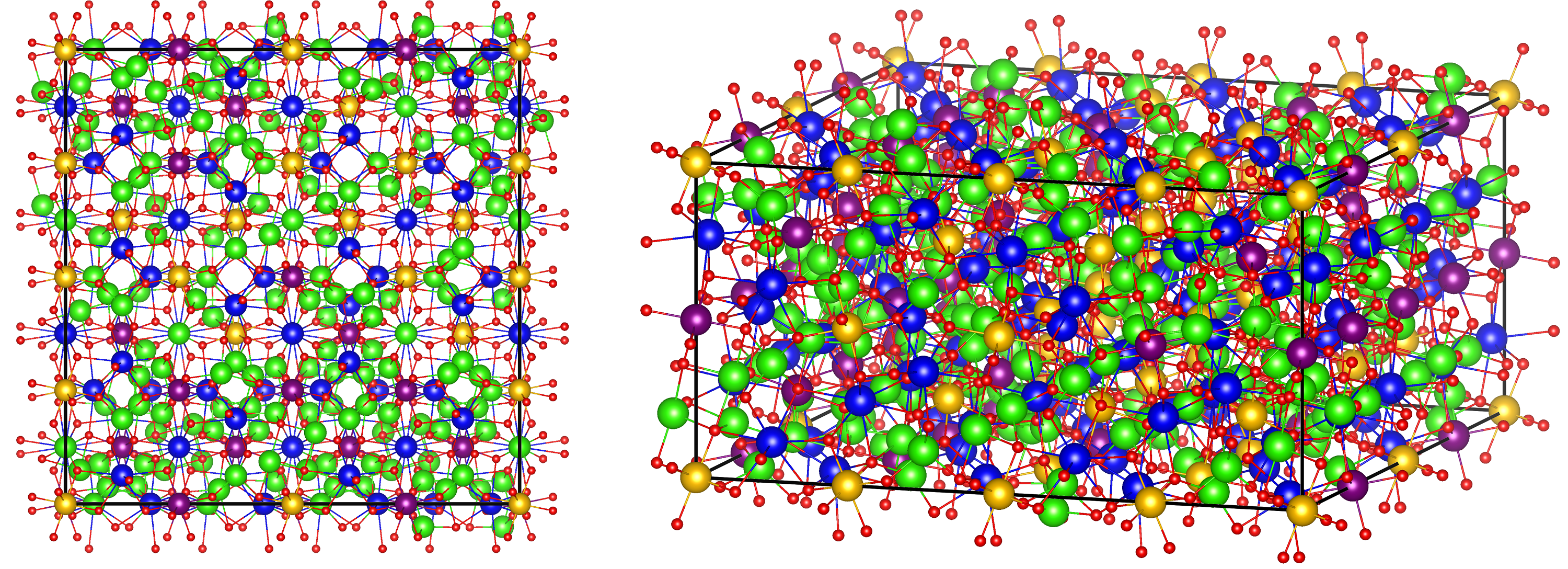}
\caption{Lowest energy structure for the optimization of the Ta-doped LLZO presented in the main text in the front (left) and perspective (right) view. The color-coding is as follows: Li: green, La: blue, Ta: purple, Zr: orange, O: red.}\label{fig:LLZO2}
\end{figure*}

\subsection*{Case Study on FeSbO\textsubscript{4}}

Yang \textit{et al.}\cite{Jang.2023} published their \textsc{EwaldSolidSolution} software very recently which allows to sample the density of states for an atomistic combinatorial problem by Coulomb energies. This technique is also able to determine minima and their code even features a local minimization algorithm similar to GD in GOAC. In their work, they solved the largest atomistic combinatorial problem by Coulomb that is, to the best of our knowledge, reported in literature. The problem is a 3$\times$3$\times$14 supercell of FeSbO\textsubscript{4} (Fe\textsuperscript{3+}, Sb\textsuperscript{5+}, O\textsuperscript{$-$2.0}) with roughly $10^{75}$ possible configurations. However, similarly to NaCl, the global minimum is still trivial and known from smaller supercells. Timings to reach the global optimum are shown for each heuristic implemented in GOAC in Table~\ref{tab:FeSbO4}.

\begin{table}[h!tb]
\caption{Average timings and statistics over 5 independent runs (each on 128 physical CPU cores, 128 OpenMP threads) with GD, MC, SA, REMC, GA, and HY as implemented in GOAC on the 3$\times$3$\times$14 FeSbO\textsubscript{4} supercell with ca. $10^{75}$ configurations. Calculations of the cluster expansion coefficient matrices, which took 155 seconds, are excluded.}\label{tab:FeSbO4}%
\begin{tabular}{@{}lcccc@{}}
\toprule
 & \multicolumn{4}{@{}c@{}}{Run Time Statistics [s]}\\
\cmidrule{2-5}%
Method & Avg. & Std. & Min. & Max.\\
\midrule
GD    & 5568 & 5877 & 3645 & 16610\\
MC    & 41 & 17 & 24 & 73\\
SA    & 17 & 2 & 15 & 20\\
REMC  & 45 & 10 & 33 & 57\\
GA\footnotemark[1]  & 100 & 74 & 29 & 199\\
HY  & 35 & 14 & 15 & 58\\
\bottomrule
\end{tabular}
\footnotetext[1]{Timings for 5 GA runs that did end in the global minimum. Some GA runs got trapped in local minima.}
\end{table}

The average timings for GD to find the global optimum are rather high and also show large deviations. The random sampling of the starting points for local minimization can be quite inefficient for this problem as there is a high probability that the algorithm ends up in a local minima. That GOAC's GD is still able to find the global minimum despite the huge computational effort spend in the many local minimization steps proves that the cluster expansion approach to the atomistic combinatorial problem is rather efficient.  

The other heuristics in GOAC are all much faster with GA being the slowest approach. It must be also noted that not all GA runs did end in the global minimum and some got trapped in local minima. This behavior might be explained by the slightly different implementation/parallelization in GA compared to the other heuristics as the parallelization is restricted to the pool creation to allow for large structure pools while for all other heuristics multiple independent runs of the same heuristic are parallelized. The SA heuristic performs best for the FeSbO\textsubscript{4} problem with the lowest average time to find the global minimum and also the lowest deviation in run time. REMC and HY also perform well on the problem but it becomes evident that their overhead in run time is larger than the actual gain in optimization performance for this problem.

It should be mentioned that the timings for GOAC correspond to the pure optimization task of the lowest energy structure and do not include any search for the density of states at higher energies. Moreover, GOAC is of heuristic nature and if the global minimum was unknown some additional run time would be required to reach a convergence criteria, e.g., $n$ heuristic steps without improvement in the global minimum, to have some certainty that the determined minimum is the global one.
\newpage

\subsection*{Structure Models and Convergence Plots for Optimization in High Entropy Layered Oxides}

\begin{figure*}[h!tb]
\centering
\includegraphics[width=0.95\textwidth]{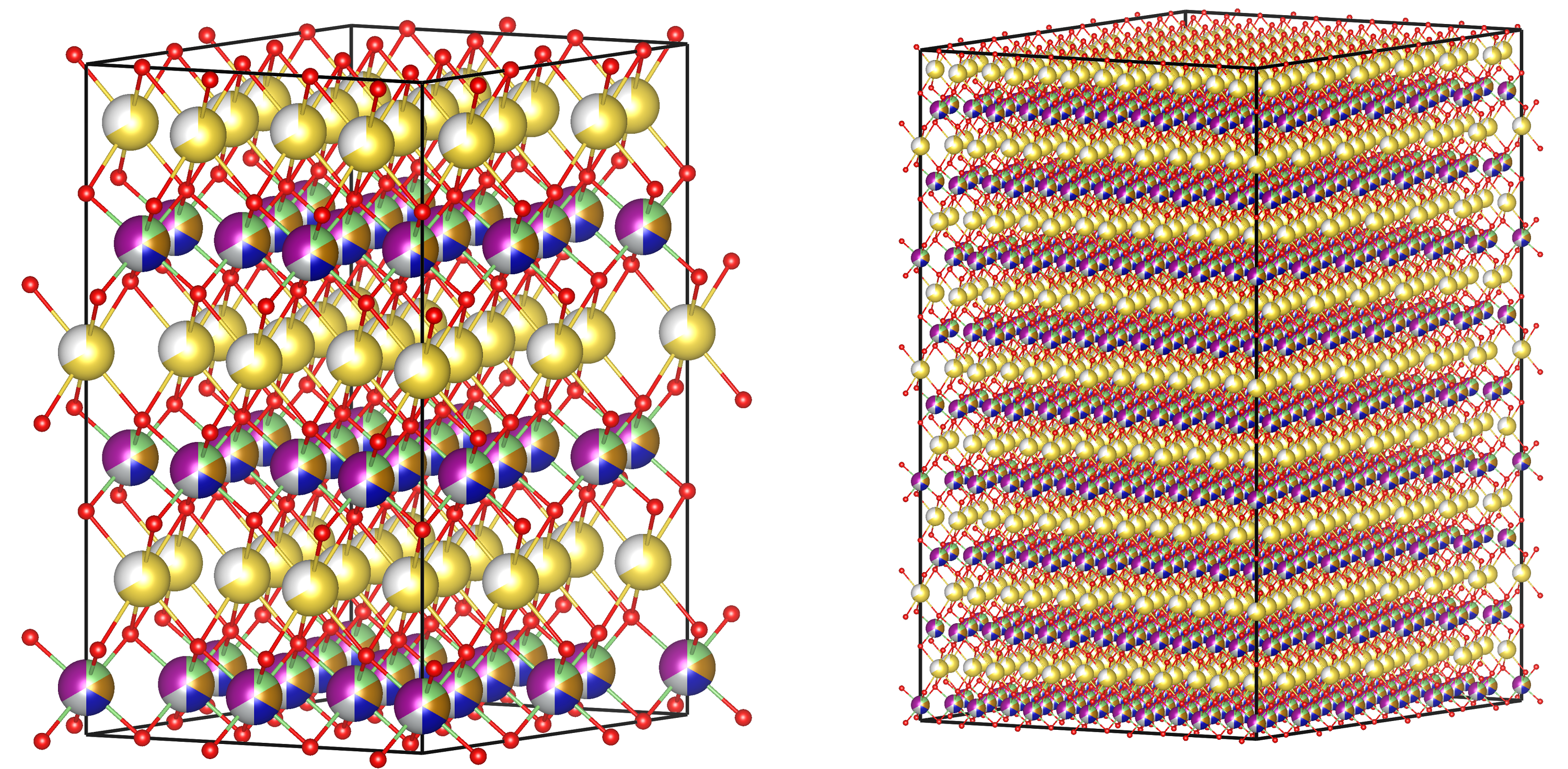}
\caption{Considered structure models of the O3-Na\textsubscript{2/3}[Li\textsubscript{1/6}Fe\textsubscript{1/6}Co\textsubscript{1/6}Ni\textsubscript{1/6}Mn\textsubscript{1/3}]O\textsubscript{2} material with partial occupations. On the left, the smallest simulated supercell (2$\times$2$\times$1) is shown and the largest one (6$\times$6$\times$3) on the right. Each structure contained just one class of Na and one class of transition-metal sites such that iteration over all shown positions in the structures was performed.}\label{fig:struct}
\end{figure*}
\newpage

\begin{figure*}[h!tb]
\centering
\includegraphics[width=0.95\textwidth]{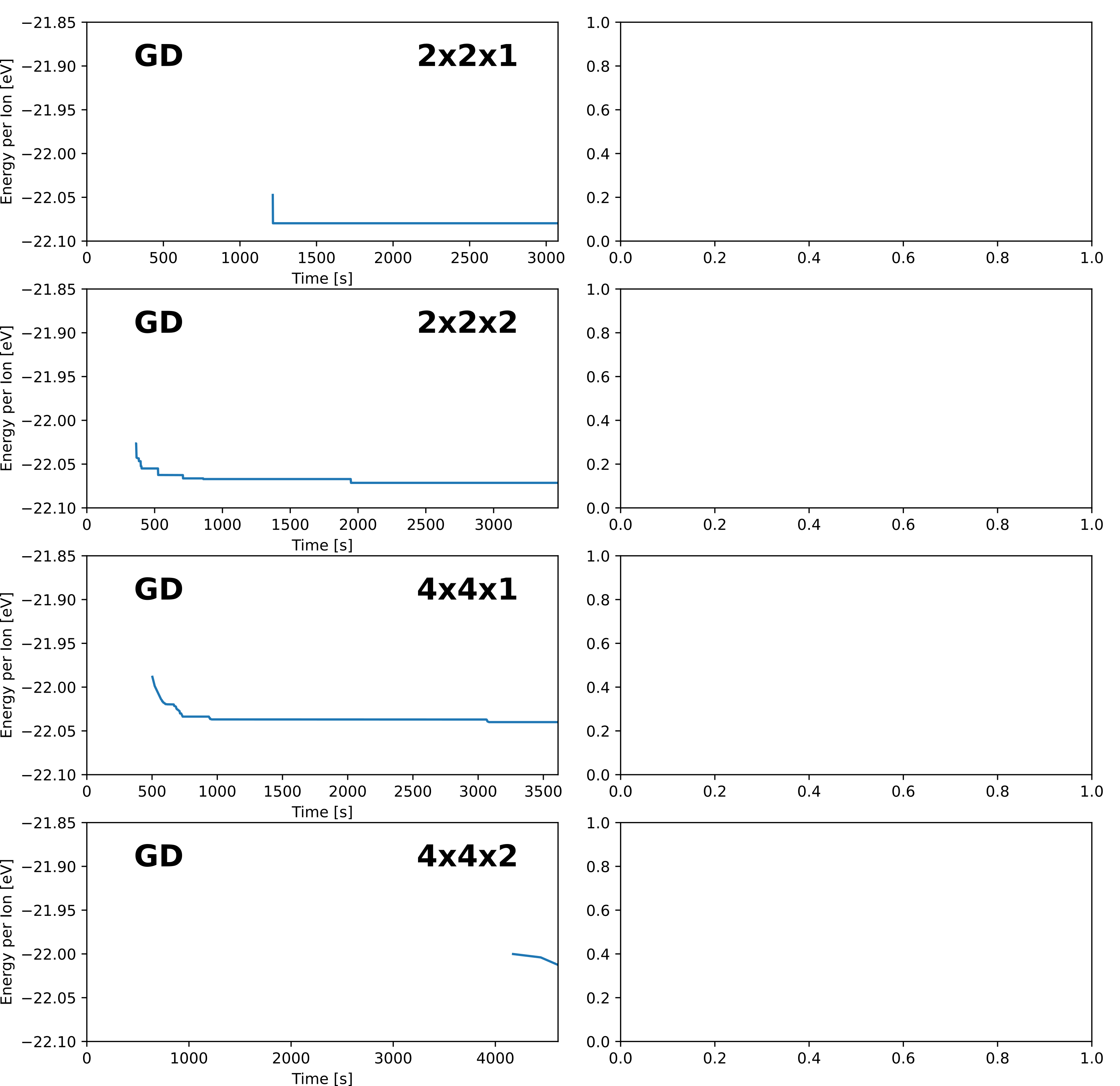}
\caption{Convergence plots for O3-Na\textsubscript{2/3}[Li\textsubscript{1/6}Fe\textsubscript{1/6}Co\textsubscript{1/6}Ni\textsubscript{1/6}Mn\textsubscript{1/3}]O\textsubscript{2} with the GD method. As discussed in the main text, only for supercells up to 4$\times$4$\times$2 solutions were obtained within the given time for this method. The first solutions start later for the 2$\times$2$\times$1 case as many more structures could be generated and treated in parallel by GD than for the more complex supercells.}\label{fig:GD}
\end{figure*}
\newpage

\begin{figure*}[h!tb]
\centering
\includegraphics[width=0.95\textwidth]{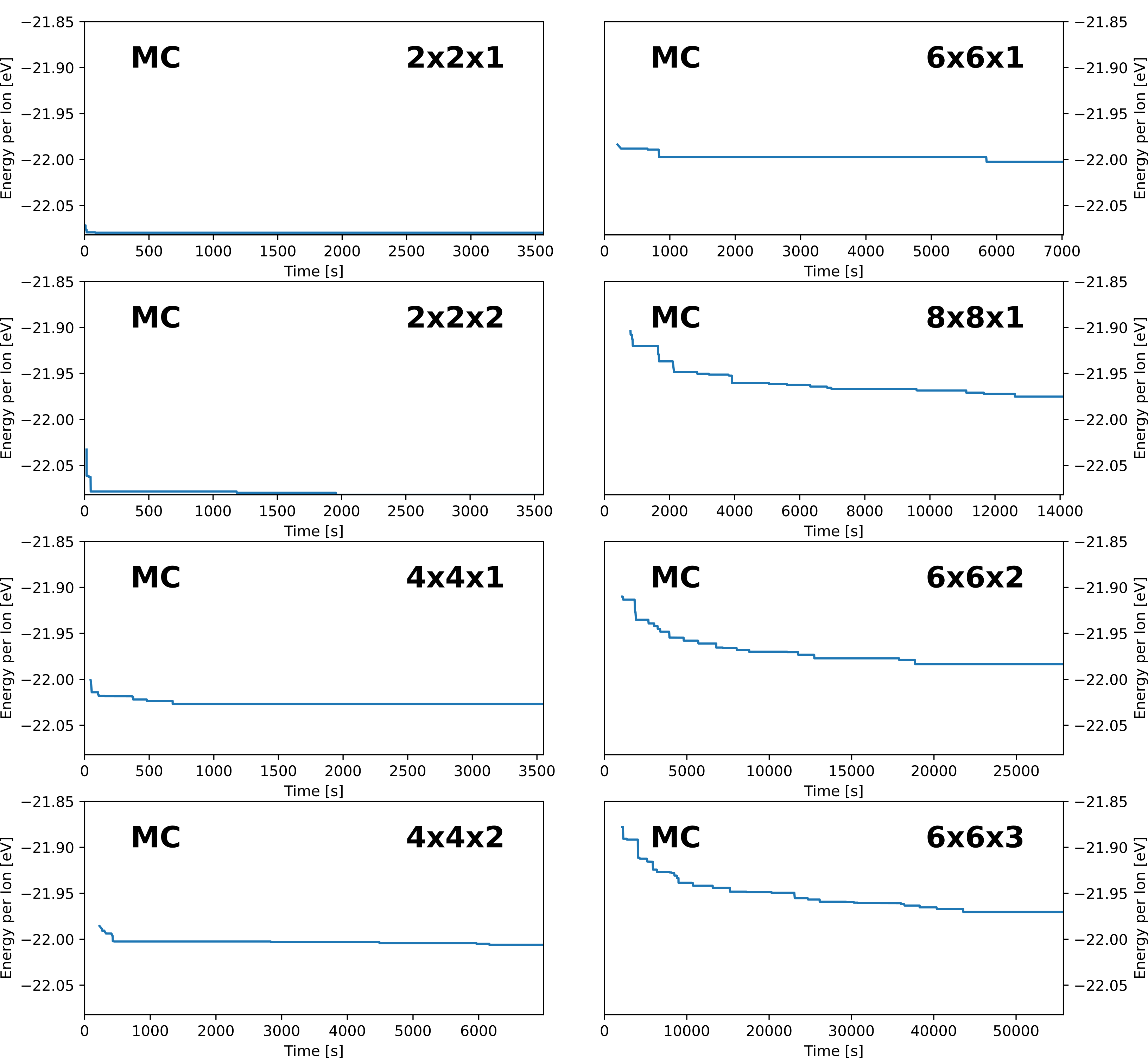}
\caption{Convergence plots for O3-Na\textsubscript{2/3}[Li\textsubscript{1/6}Fe\textsubscript{1/6}Co\textsubscript{1/6}Ni\textsubscript{1/6}Mn\textsubscript{1/3}]O\textsubscript{2} with the MC method. The plots nicely show that for more easy problems (small supercells) the best solution is almost found instantly, while in larger supercells a slow convergence over run time can be observed.}\label{fig:MC}
\end{figure*}
\newpage

\begin{figure*}[h!tb]
\centering
\includegraphics[width=0.95\textwidth]{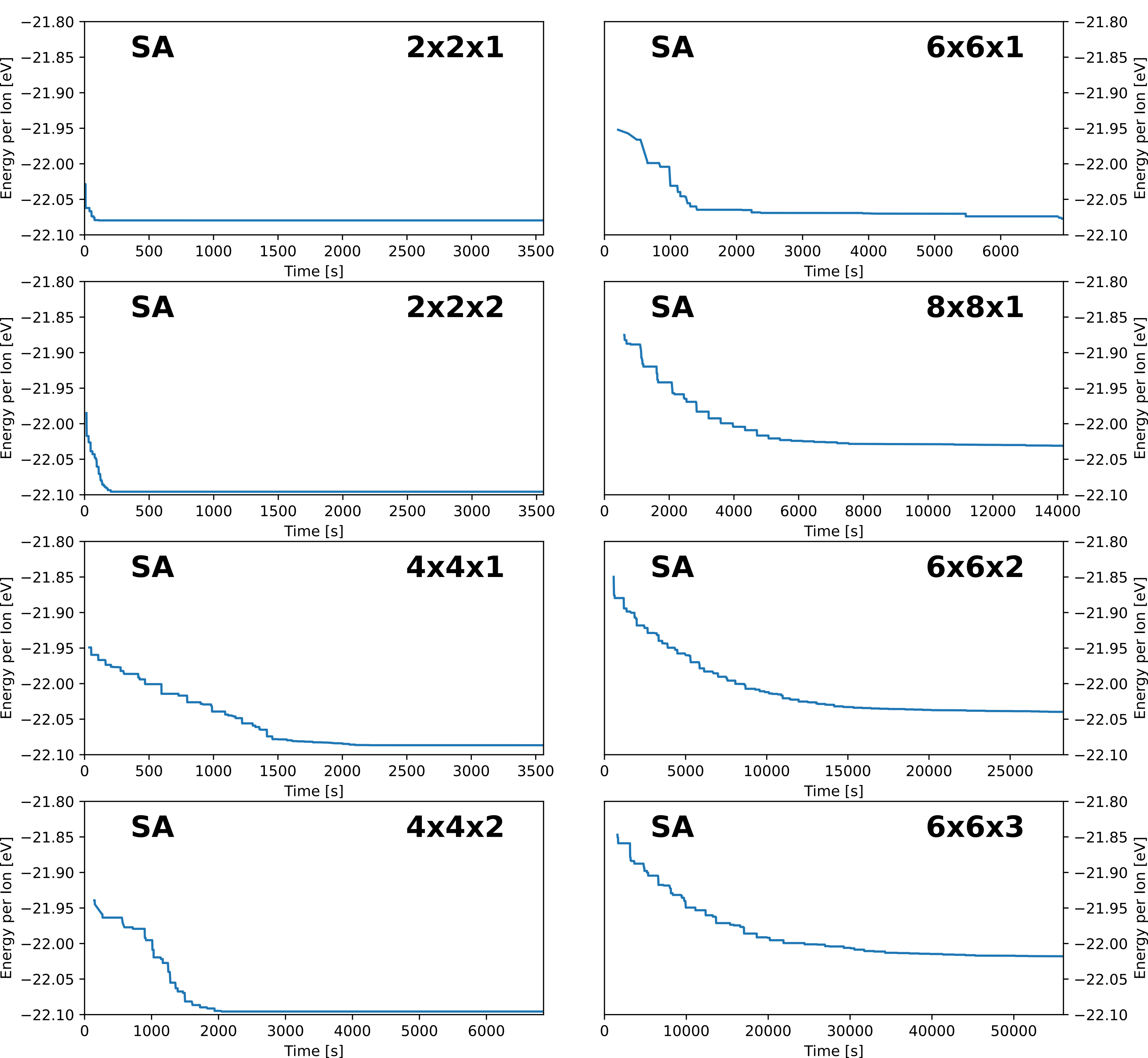}
\caption{Convergence plots for O3-Na\textsubscript{2/3}[Li\textsubscript{1/6}Fe\textsubscript{1/6}Co\textsubscript{1/6}Ni\textsubscript{1/6}Mn\textsubscript{1/3}]O\textsubscript{2} with the SA method. The convergence trends are similar to the pure MC method but convergence over time is more pronounced and even looks like an exponential decay for larger supercells. This is in good agreement with the temperature evolution of the SA method that also decreases the temperature exponentially over run time.}\label{fig:SA}
\end{figure*}
\newpage

\begin{figure*}[h!tb]
\centering
\includegraphics[width=0.95\textwidth]{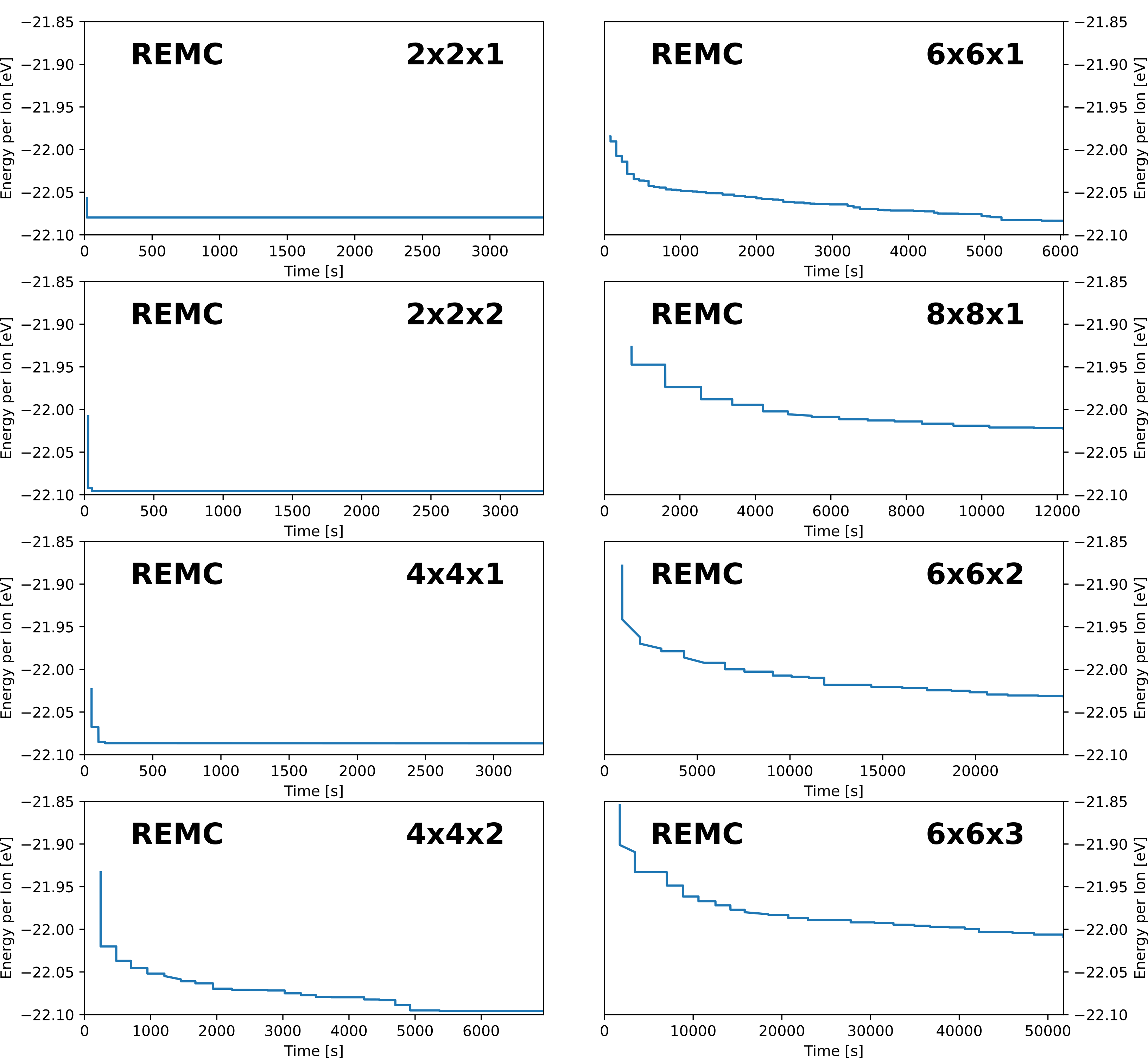}
\caption{Convergence plots for O3-Na\textsubscript{2/3}[Li\textsubscript{1/6}Fe\textsubscript{1/6}Co\textsubscript{1/6}Ni\textsubscript{1/6}Mn\textsubscript{1/3}]O\textsubscript{2} with the REMC method. Also REMC converges almost instantly in the smaller supercells and shows almost in all examples a steep decrease in energy within the first seconds of run time.}\label{fig:REMC}
\end{figure*}
\newpage

\begin{figure*}[h!tb]
\centering
\includegraphics[width=0.95\textwidth]{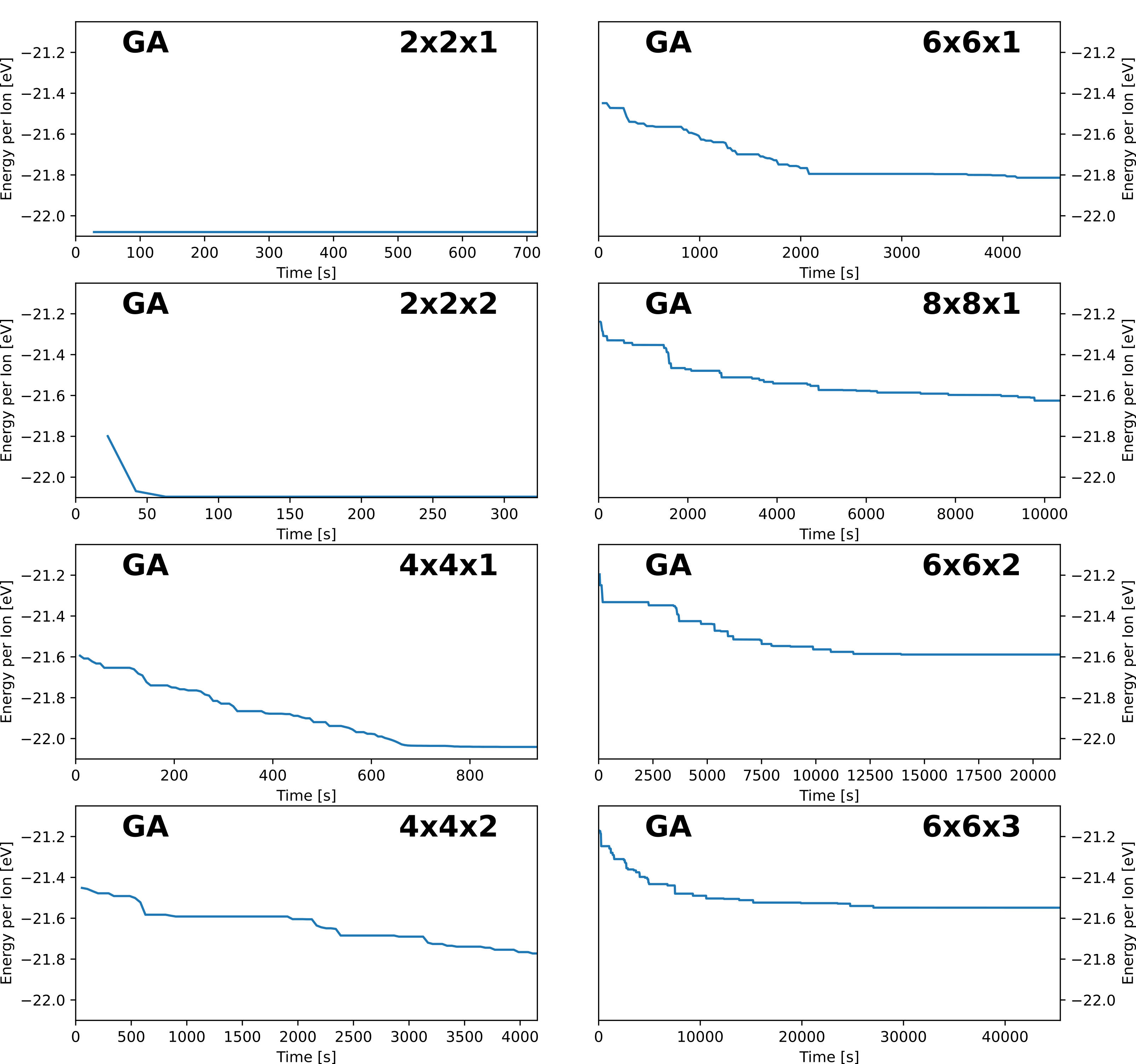}
\caption{Convergence plots for O3-Na\textsubscript{2/3}[Li\textsubscript{1/6}Fe\textsubscript{1/6}Co\textsubscript{1/6}Ni\textsubscript{1/6}Mn\textsubscript{1/3}]O\textsubscript{2} with the GA method. The GA also converges within almost no run time for the small supercells. While for intermediate sized supercells (4$\times$4$\times$1 to 8$\times$8$\times$1) a steady, even though slowing down by supercell size, convergence is observed. For even larger supercells the GA gets trapped in one solution after roughly half of the run time. This highlights that the GA is prone to get trapped in local minima for complex problems.}\label{fig:GA}
\end{figure*}
\newpage

\begin{figure*}[h!tb]
\centering
\includegraphics[width=0.95\textwidth]{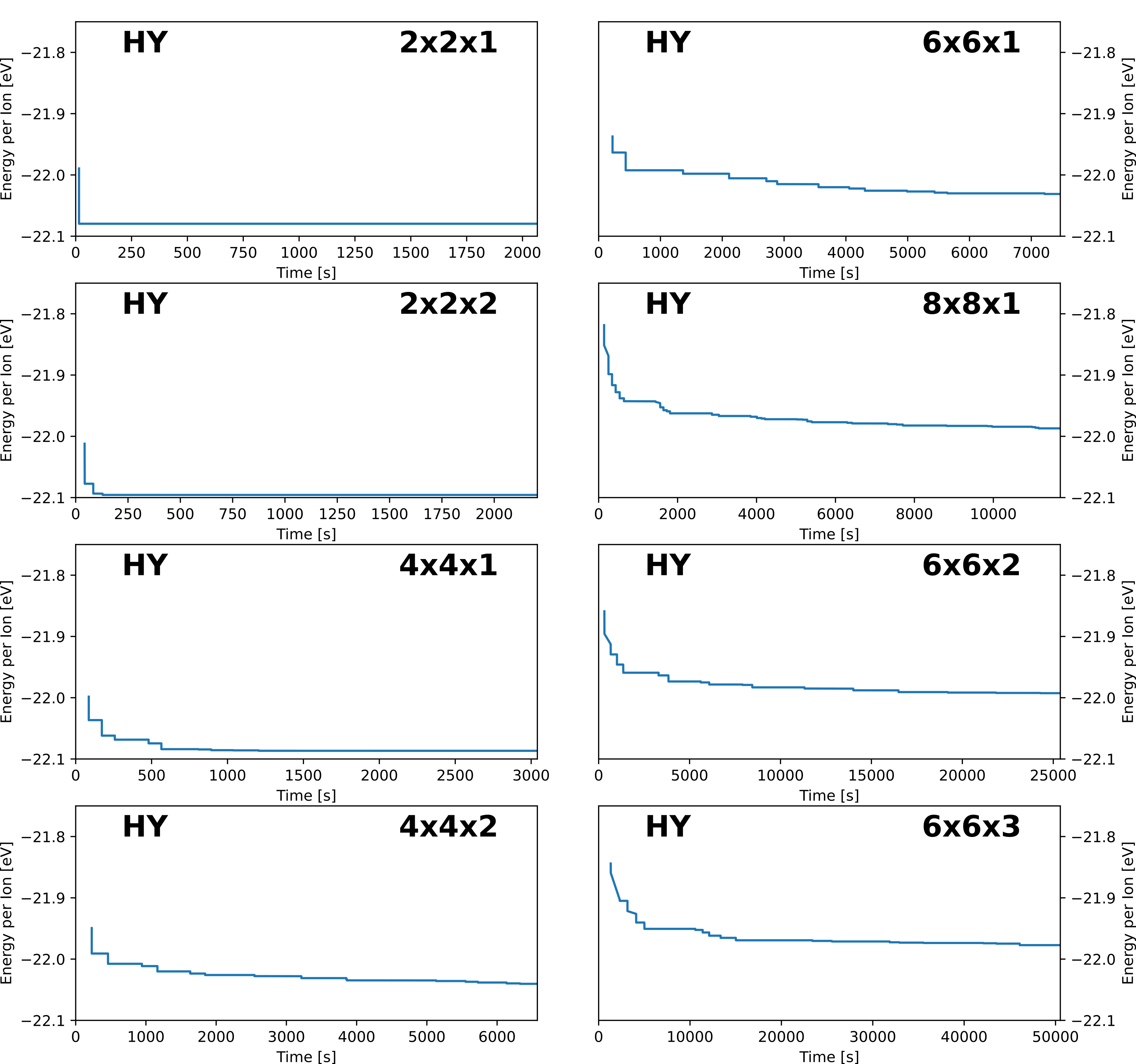}
\caption{Convergence plots for O3-Na\textsubscript{2/3}[Li\textsubscript{1/6}Fe\textsubscript{1/6}Co\textsubscript{1/6}Ni\textsubscript{1/6}Mn\textsubscript{1/3}]O\textsubscript{2} with the HY (REMC + GA) method. For the small supercells similar trends than for the pure methods can be observed. At larger supercells, however, the convergence is slowed down compared to the pure REMC method. In contrast to the pure GA, even after longer run times small improvements in energy are observed. For the most complex problems it appears that the HY approach is only beneficial for the first seconds/minutes while at longer run times the pure REMC method clearly outperforms the HY approach.}\label{fig:HY}
\end{figure*}

\end{document}